\documentclass[lettersize,journal]{IEEEtran}
\usepackage{amsmath,amsfonts}
\usepackage{algorithmic}
\usepackage{algorithm}
\usepackage{array}
\usepackage[caption=false,font=normalsize,labelfont=sf,textfont=sf]{subfig}
\usepackage{textcomp}
\usepackage{stfloats}
\usepackage{url}
\usepackage{verbatim}
\usepackage{graphicx}
\usepackage{cite}
\usepackage{enumerate}
\usepackage{bm}
\usepackage{tikz}
\usepackage{comment}
\usepackage{amsmath,amssymb} 
\usepackage{color}
\usepackage{multirow}
\usepackage{tabularx}
\usepackage[utf8]{inputenc}
\usepackage{booktabs} 

\usepackage{color}

\begin{document}

\title{ChiMera: Learning with noisy labels by contrasting mixed-up augmentations}

\author{Zixuan Liu*,~\IEEEmembership{Student Member,~IEEE}, Xin Zhang*, Junjun He, Dan Fu, Dimitris Samaras, Robby Tan, Xiao Wang, Sheng Wang
\thanks{Zixuan Liu and Xin Zhang have equal contribution.}
}

\markboth{IEEE TRANSACTIONS ON Pattern Analysis and Machine Intelligence ,~Vol.~14, No.~8, May~2023}%
{Shell \MakeLowercase{\textit{et al.}}: A Sample Article Using IEEEtran.cls for IEEE Journals}


\maketitle

\begin{abstract}
Learning with noisy labels has been studied to address incorrect label annotations in real-world applications. In this paper, we present ChiMera, a two-stage learning-from-noisy-labels framework based on semi-supervised learning, developed based on a novel contrastive learning technique MixCLR. The key idea of MixCLR is to learn and refine the representations of mixed augmentations from two different images to better resist label noise. ChiMera jointly learns the representations of the original data distribution and mixed-up data distribution via MixCLR, introducing many additional augmented samples to fill in the gap between different classes. This results in a more smoothed representation space learned by contrastive learning with better alignment and a more robust decision boundary. By exploiting MixCLR, ChiMera also improves the label diffusion process in the semi-supervised noise recovery stage and further boosts its ability to diffuse correct label information. We evaluated ChiMera on seven real-world datasets and obtained state-of-the-art performance on both symmetric noise and asymmetric noise. Our method opens up new avenues for using contrastive learning on learning with noisy labels and we envision MixCLR to be broadly applicable to other applications.
\end{abstract}


\section{Introduction}
\IEEEPARstart{R}{eal}-world classification problems often present substantial noise in the annotated labels. Deep neural networks are prone to overfitting on data with noisy labels, resulting in substantial prediction performance drop \cite{zhang2016understanding}. As a result, learning with noisy labels (LNL) has been extensively studied to perform supervised learning on data with noisy labels. Most existing LNL approaches can be divided into two categories: identifying and correcting wrong labels \cite{li2020dividemix,zhang2020decoupling,han2019deep,tanaka2018joint,yi2019probabilistic, li2022selec,zhao2022centrality, li2023disc} or regularizing classification loss \cite{patrini2017making,goldberger2016training,liu2020early,bai2021understanding,englesson2021generalized, bae2022noisy, yan2022noise,yi2022learning}. Identifying clean labels is an important and efficient way to utilize label information, as the tractability of LNL problems comes from the assumption that the population of samples with clean labels is statistically larger than samples with corrupted labels. Such a process can be addressed by performing noise detection to specify the major sample cluster and drop the minor clusters \cite{li2020dividemix, zhao2022centrality}. 
To correct wrong labels, a streamlined solution has two steps:
(i) first learn good sample representations and a similarity measure by extracting and comparing critical patterns from samples (representation learning), 
(ii) and then learn to generalize the correct label information from clean samples to noisy samples with similar patterns (label diffusion). 
The key to this solution is representation learning. Because the sample representations and the corresponding similarity measures should be both helpful to noise detection and label diffusion.


\begin{figure}[ht]
\centering
 \includegraphics[width=\linewidth]{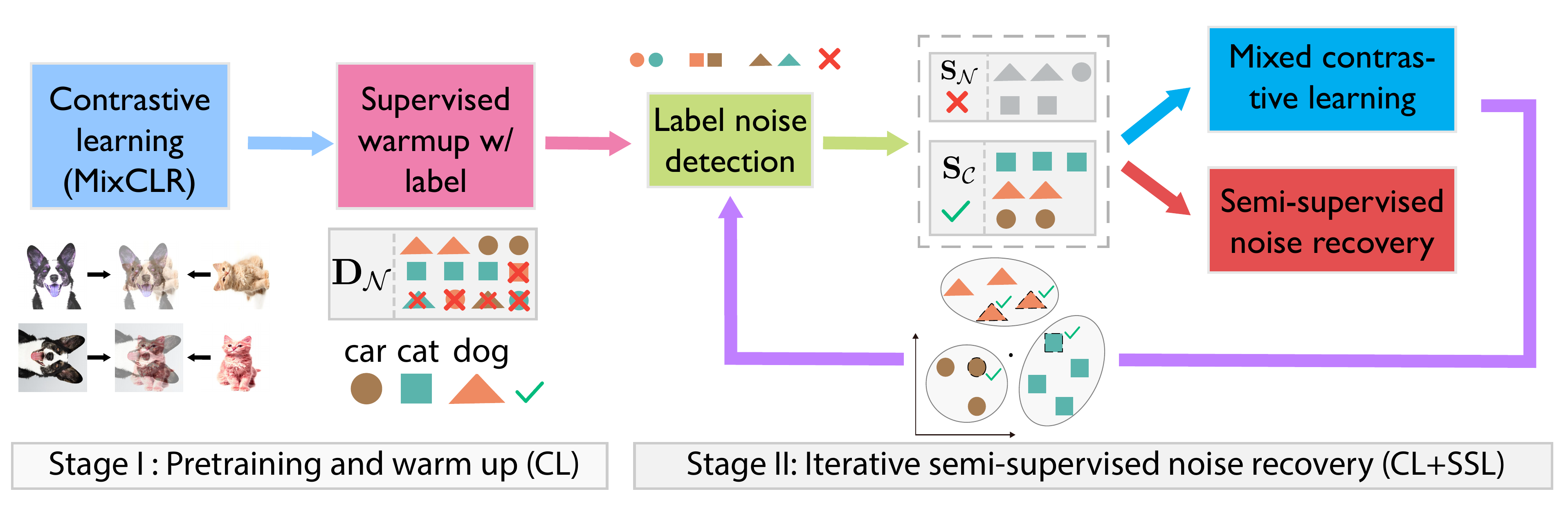}
  \setlength{\abovecaptionskip}{-0.1cm}
\caption{Overview of ChiMera. It is a unified two-stage framework to address the learning with noisy labels problem by leveraging contrastive representation learning and semi-supervised learning. The first stage is a pretraining warm-up stage and the second stage is a semi-supervised noise recovery stage.
}
\label{fig:figure1_simplified}
\end{figure}

Contrastive Learning (CL) has recently shown encouraging results in learning good representations by projecting augmented views close to each other in the embedding space (Fig. \ref{fig:alignment_and_tsne}a) \cite{chen2020simple,he2020momentum,chen2020big,chen2020improved}. CL has also been exploited in learning with noisy labels through learning class prototypes \cite{li2021learning, yi2022learning}, improving network initialization \cite{zheltonozhskii2021contrast}, or regularizing classification losses \cite{yi2022learning}. Previous work has explained the superior performance of CL using its improved uniformity and alignment on the hypersphere \cite{wang2020understanding,zimmermann2021contrastive}. However, features learned from these methods might not be robust enough to handle noisy labels in the classification stage, since noisy labels might still provide wrong supervision that will distort the learned representations \cite{li2022selec}. We therefore hypothesize that learning a classification hyperplane on this hypersphere is still sensitive to noisy labels, thus making CL less applicable to LNL. Intuitively, augmenting additional data points in this hypersphere might help improve alignment and uniformity, leading the learned hyperplane to be less sensitive to noisy labels. 

\begin{figure}[ht]
\centering
 \includegraphics[width=\linewidth]{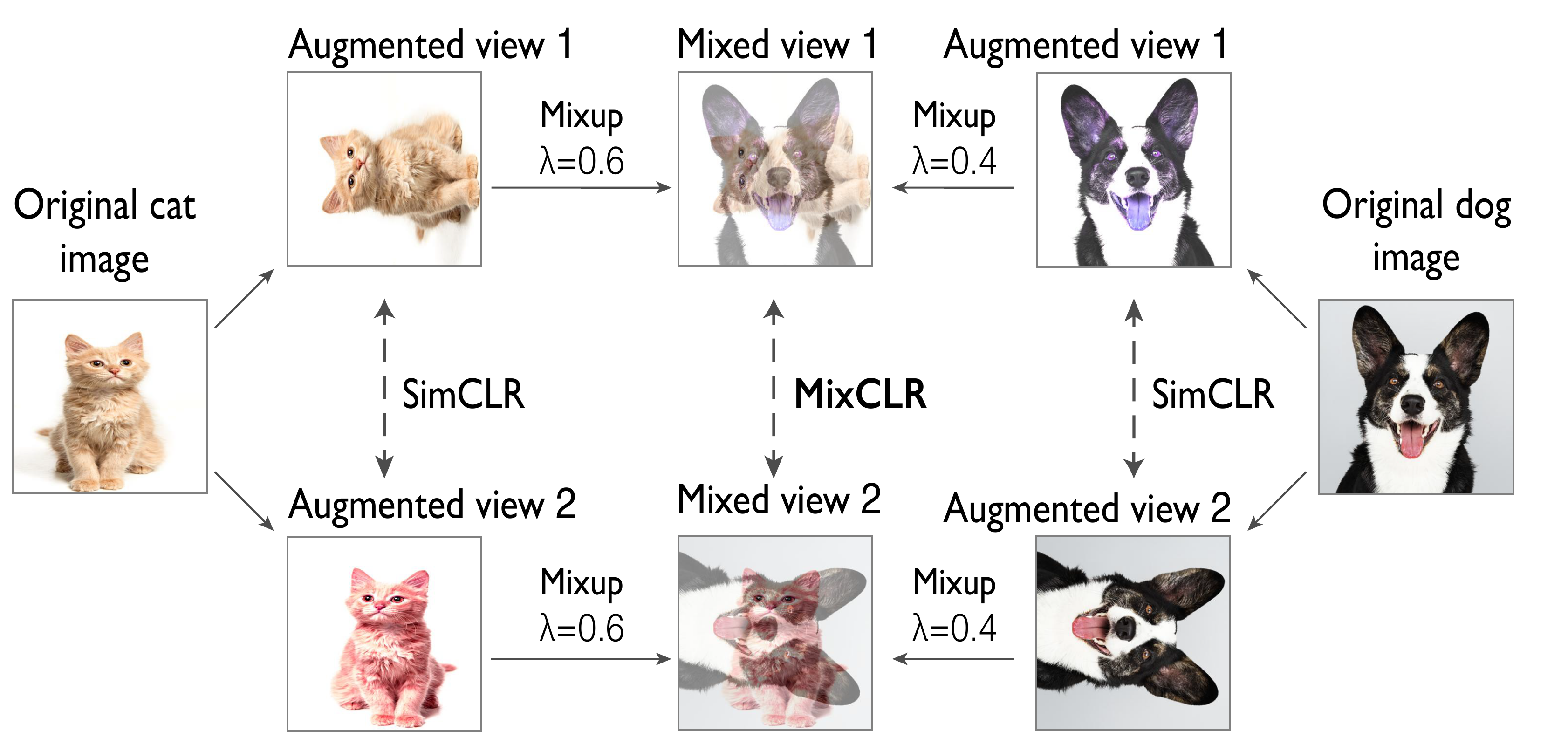}
  \setlength{\abovecaptionskip}{-0.1cm}
\caption{The key idea of MixCLR is to construct mixed positive pairs and apply contrastive learning to them. The representations of the positive pair will be close to each other as well as their seed images, resulting in a smoother representation space.
}
\label{fig:mixclr_illustration}
\end{figure}

In this work, we present a novel LNL framework ChiMera (Fig. \ref{fig:chimera_flowchart}) that is agnostic to label noise type (e.g., random perturbed noise or asymmetric noise) following this intuition. ChiMera is a two-stage detect-then-correct framework that iteratively drops labels that are possibly noisy and utilizes semi-supervised learning to update. The key idea of ChiMera is to learn and refine the representations of mixed augmentations of two different samples by performing both contrastive learning and semi-supervised learning. Each positive pair of samples is created by mixing augmentations from two images through mixup augmentations and then learned via contrastive learning. We name this novel contrastive learning technique MixCLR (Fig. \ref{fig:mixclr_illustration}) and apply it to both the pretraining stage and the semi-supervised learning stage in our framework. We find improved performance under label noise resulting from two key advantages of MixCLR - (i) it helps to learn a representation space with better alignment and robust decision boundary, and (ii) it improves the label diffusion by jointly learning original data distribution and mixed-up data distribution via selective-supervision signals. Furthermore, we propose AsyMixCLR to deal with asymmetric noisy labels, which mixes samples from classes with similar appearances for better discrimination. Though ChiMera is agnostic to different label types, AsyMixCLR enlarges its ability to mine and utilize priors hidden in the dataset. 

We conducted extensive experiments on seven datasets with three different types of simulated noise and real-world noise, observing substantial improvement of our method over the state-of-the-art approaches on both symmetric, asymmetric label noise as well as instance-dependent label noise. 
ChiMera improves at least 0.94\% and 3.65\% on symmetric label noise tasks on Cifar-10 and Cifar-100. Furthermore, ChiMera achieves from 0.77\% to 1.79\% on 6 variants of CIFAR-10 and CIFAR-100 with real-world noise injected.
Finally, ChiMera is shown to be effective in resisting simulated and real-world instance-dependent and asymmetric noise from 0.5\% to 1.5\%.
Our main contributions are:
\begin{enumerate}
    \item We present a novel learning with noisy labels framework ChiMera, which can effectively resist label noise via semi-supervised learning and contrastive learning. 
    \item We investigate how contrastive learning can be boosted by leveraging self-supervision on mixed samples and propose MixCLR to improve both the weakly-supervised pretraining and semi-supervised noise recovery.
    \item We extend MixCLR to AsyMixCLR to effectively address asymmetric label noise.
    \item We deliver extensive ablation studies to fully verify the effectiveness of ChiMera and MixCLR.
\end{enumerate}

\section{Related works}

\subsection{Learning with Noisy Label}

\label{sec:related_lnl}
Many studies focus on reducing the effect of label noise and generalizing the correct labels. These methods can be mainly divided into three categories: regularizing the noise, sample selection and reutilizing noisy negatives, and learning from extra clean validation signal.
Methods belonging to the first type explore ways to adjust classification loss by estimating noise transition matrix \cite{patrini2017making,goldberger2016training,cheng2022instance,yang2022estimating,bae2022noisy,jiang2022information}, re-weighting samples~\cite{zhang2023learning} by designing criterions~\cite{wang2022scalable} such as small-loss \cite{jiang2018mentornet,han2018co,englesson2021generalized} and prediction disagreement \cite{malach2017decoupling}, or directly applying regularization~\cite{liu2022robust,cheng2022class,iscen2022learning} through early-stop strategy \cite{liu2020early,bai2021understanding}. 
However, these methods may depend on the noise distribution and do not leverage the full potential of the samples with noisy labels. In contrast, our method exploits contrastive learning and semi-supervised learning to detect and learn from samples with noisy labels step by step, thus can better handle label noise. 
The second type of methods started by detecting the wrong labels and recently further attempted to correct them~\cite{tu2023learning,li2023disc} by learning class prototypes \cite{han2019deep}, predicting pseudo labels, treating labels as learnable latent variables \cite{tanaka2018joint,yi2019probabilistic,sun2022pnp}. Among these methods, noise reduction methods via sample selection \cite{xia2021sample,cheng2020learning,feng2023ot,zhao2022centrality,li2023disc, xiao2023promix} have been shown effectively to have better noise correction ability. Some of these methods selected samples based on confidence score \cite{bai2021me} or other metrics aggregated from model predictions \cite{cordeiro2021longremix}. 

Another promising direction which our method ChiMera also can be clustered into is to combine noise detection module with semi-supervised learning 
\cite{li2020dividemix,nishi2021augmentation,cordeiro2021longremix,bai2021understanding}. 
These methods principally do not require understanding knowledge of the noise distribution and largely rely on the recent success of semi-supervised learning \cite{berthelot2019mixmatch, berthelot2019remixmatch}. Therefore, their performance will degrade when the noise ratio is high and clean supervision is hard to mine or the noise distribution is complex \cite{zhang2022learning}. There are also attempts to exploit an extra validation set to provide clean supervision \cite{bai2021me, tu2023learning} to help further purify the noisy label and have achieved promising performance. However, this requires extra clean supervision, which is not always accessible. 
A few other recent studies have explored new metrics \cite{Zhu_2021_CVPR} or new noise types such as instance-dependent noise \cite{berthon2021confidence,xia2021sample,bai2021me,cheng2020learning,Zhu_2021_CVPR,zhao2022centrality}, and new settings of noisy label learning \cite{wu2021ngc,li2021learning}, however, they are either noise-type specific or constraint to extra assumptions. There are also some approaches that leverage contrastive learning to acquire prototypes \cite{li2021learning, zhao2022centrality} or robust representations  \cite{zhang2020decoupling,zheltonozhskii2021contrast,xue2022investigating,huang2023twin,yan2022noise}, but these methods either highly rely on the learned representations and use simplified noise reduction design, or slightly adapt existing methods to incorporate contrastive learning, which not fully leverage the full potential of contrastive learning and advanced understanding gained from existing LNL methods. Our method ChiMera proposes to harmoniously combine contrastive learning and semi-supervised learning together via a novel contrastive technique MixCLR, leading to superior performance on resisting multiple types of label noise.

\subsection{Contrastive Learning}
Contrastive learning methods directly regularize the representation space by encouraging representations of different augmented views from the same images together and spreading views from different images apart via InfoNCE loss \cite{oord2018representation}. 
Several popular realizations \cite{chen2020simple,he2020momentum,grill2020bootstrap,chen2020exploring} of contrastive learning framework have empirically shown that their learned representations result in substantial improvements in the downstream classification tasks. The representations learned by contrastive learning have shown promising performance in pattern extraction and clustering. Besides the success achieved by self-supervised contrastive learning, the supervised contrastive learning framework \cite{khosla2020supervised}, which treats images from the same classes as positive pairs, also achieves comparable or even better performance than widely-used classification frameworks based on conventional cross-entropy loss.

To explain why it achieves great empirical success, many efforts seek to understand why contrastive learning works. \cite{wang2020understanding} pointed out that contrastive learning asymptotically optimizes for alignment and uniformity properties and proposed two quantifiable metrics to measure the quality of representations. \cite{zimmermann2021contrastive} introduced that contrastive learning can be treated as inverting the data generation process on observed datasets. \cite{aitchison2021infonce} stated that contrastive learning can be viewed as training a self-supervised variational autoencoder. 

To improve the robustness of the learned representations, one simple yet effective approach is to expand the dataset via mixup augmentation \cite{zhang2017mixup}. 
The idea of mixup has been proved to be robust to adversarial data noise \cite{zhang2020does} and achieves empirical success on the semi-supervised learning \cite{berthelot2019mixmatch,berthelot2019remixmatch,sohn2020fixmatch}.
Since applying contrastive learning requires us to construct augmented views, there are a few attempts to improve contrastive learning performance by leveraging mixup augmentation. DACL \cite{verma2021towards} tries to anchor contrastive learning on raw images and utilize mixup as an extra random augmentation. Both Mix-Co \cite{kim2020mixco} and Un-Mix \cite{shen2020mix} construct positive pairs anchored on mixed-up images, but instead consider the two template images as the positive samples with re-weighted importance. Since the encoder is a complex non-convex function, this could be problematic when $\lambda$ is far away from 0 or 1. In contrast, our proposed MixCLR performs contrastive learning on mixed-up augmentations and thus does not need to calculate the distance between template samples.  

\section{Preliminary and notation}
In this section, we provide the necessary background and several notations that will be used later for learning with noisy labels and contrastive learning, and discuss the challenges of classifier-based
noise recovery and contrastive learning. 
\subsection{Learning with Noisy Labels (LNL)}
\label{sec:pre_lnl}
\subsubsection{Problem setting and feasibility}
\label{sec:pre_lnl_ps}
Let $\mathbb{D}_\mathbf{x}$ be the the raw data space, $\mathcal{D}_\mathbf{x}$ be a distribution on $\mathbb{D}_\mathbf{x}$ and $C$ be the number of classes, the aim of LNL is to solve a classification task $\mathbb{D}_\mathbf{x} \rightarrow \{0,1\}^C$ with a noisy labeled training dataset $\mathbf{D}_{\mathcal{N}} = \{(\mathbf{x}_i,\mathbf{\tilde{y}}_i)\}_{i=1}^{|\mathbf{D}|}$ and achieve good performance on a correctly labeled test set (we refer to as `clean' in the following) $\mathbf{D}_\mathcal{T} = \{(\mathbf{t}_i,\mathbf{s}_i)\}_{i=1}^{|\mathbf{D}_\mathcal{T}|}$, where $\mathbf{x}_i,\mathbf{t}_i \sim \mathcal{D}_{\mathbf{x}} \in \mathbb{D}_\mathbf{x}$ and $\mathbf{\tilde{y}}_i,\mathbf{s}_i \in \{0,1\}^C$. In other words, it is equivalent to estimating a good $P(\mathbf{y}|\mathbf{x})$ solely based on $\mathbf{D}_{\mathcal{N}}$. Here, however, the correct label $\mathbf{y}_i \in \{0,1\}^{C}$ of the training sample $\mathbf{x}_i$ is unknown, that is, it is unknown whether $\mathbf{\tilde{y}}_i = \mathbf{y}_i$ is true for any $\mathbf{x}_i \in \mathbf{D}_{\mathcal{N}}$. 

The above setting is general as it only assumes the existence of noisy labels, but does not assume the scale of label noise and why they exist. Despite not knowing which of the training samples are noisy, the assumption that only a relatively small group (not the major) of the training data is wrongly annotated is required. Specifically, let $\mathbf{\tilde{C}} = \sum^{|\mathbf{D}_\mathcal{T}|}_{i=1} \mathbf{y}_i\mathbf{\tilde{y}}_i^T \in \mathbf{R}^{C \times C}$ be the label-noise confusion matrix of $\mathbf{D}_{\mathcal{N}}$ with entries \( \tilde{c}_{jk} \) for row \( j \) and column \( k \), a feasible learning with noisy label task requires $\tilde{c}_{jj} = \max_k \tilde{c}_{jk}, 1\leq k \leq C$ for all $1 \leq j \leq C$ to have enough correct labeled training samples for each class. 
\begin{figure*}[ht]
\centering
 \includegraphics[width=\linewidth]{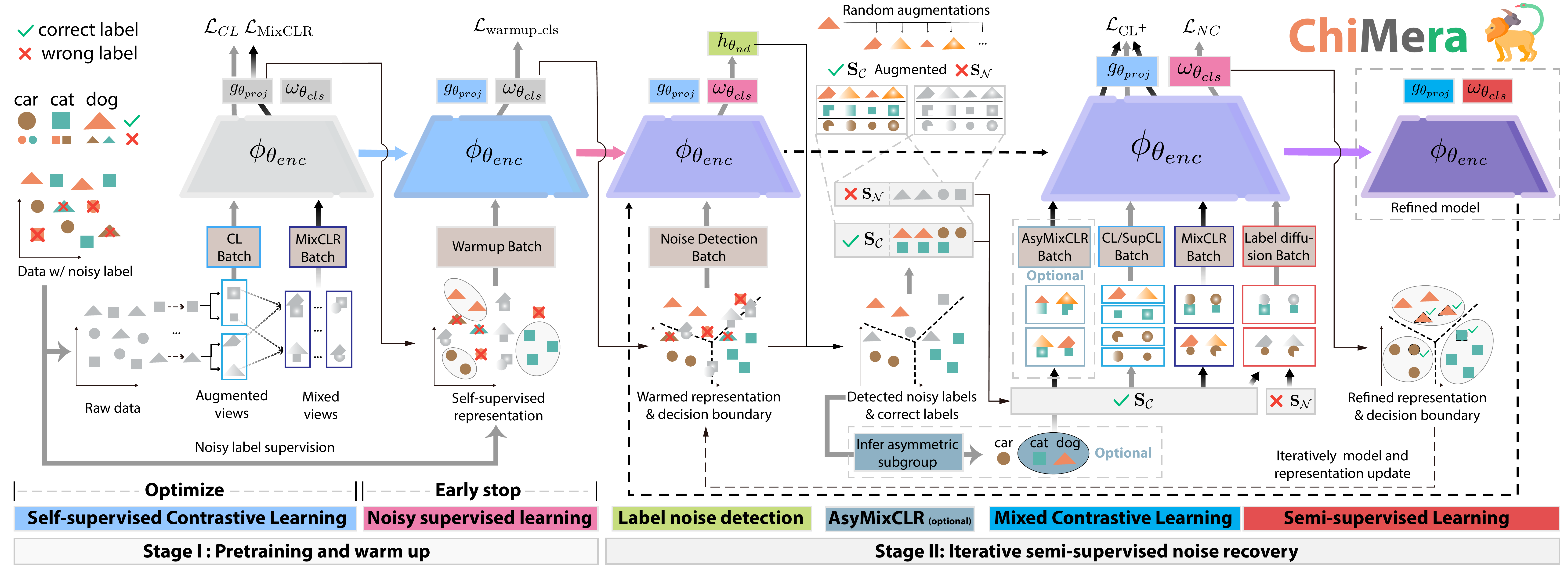}
  \setlength{\abovecaptionskip}{-0.1cm}
  \setlength{\belowcaptionskip}{-0.1cm}
\caption{The detailed workflow of ChiMera. MixCLR is crucial in both stages. In the first stage, MixCLR helps ChiMera to learn robust representation and resist noisy supervised warm-up. In the second stage, the noise detector splits the noisy dataset into clean and noisy subsets, and MixCLR and mixup-based semi-supervised learning are used to further reduce the noise and improve the model.
}
\label{fig:chimera_flowchart}
\end{figure*}
\subsubsection{Probablistic label noise recovery modeling}
\label{sec:pre_lnl_nt}
In reality, different types of noisy labels can occur and are hard to quantify due to many reasons, such as unexpected label matching errors, imprecise annotation, and challenging ambiguous samples. However, since estimating a good $P(y|x)$ should also achieve good performance on $\mathbf{D}_{\mathcal{N}}$, solving LNL can also be seen as recovering the perturbed label noise to its ground-truth. 
We can thus model most of the noise label recovery process with the following probabilistic noise recovery model:    
\begin{equation}
     P(\mathbf{y}_i|\mathbf{\tilde{y}}_i, \mathbf{x}_i) =\left\{
\begin{aligned}
r_i& , &  \mathbf{y}_i \sim \mathcal{P}_\mathcal{N}(\mathbf{x}_i, \tilde{\mathbf{y}}_i), \\
1-r_i & , & \mathbf{y}_i=\tilde{\mathbf{y}}_i.
\end{aligned}
\right.
\label{eqn:pre_noisechannel}
\end{equation}

Here $r_i$ is the instance-dependent label flipping ratio and  $\mathcal{P}_\mathcal{N}(\mathbf{x}, \tilde{\mathbf{y}})$ is the instance-dependent probability distribution over all classes except the correct class. 
Suppose the ground truth labels for all training instances are known, the weighted confusion matrix $\mathbf{\overline{C}} = \mathbf{\tilde{C}} / |\mathbf{D}_{\mathcal{N}|}$ with entries $\overline{c}_{jk}$ for row \( j \) and column \( k \) can be seen as the empirical noise transition matrix of $\mathbf{D}_{\mathcal{N}}$. 
Nonetheless, it is difficult to estimate $r_i$ and $\mathcal{P}_\mathcal{N}(\mathbf{x}, \tilde{\mathbf{y}})$ at the instance level solely based on $\mathbf{\overline{C}}$. An easier way to leverage $\mathbf{\overline{C}}$ is to acquire an estimation of class-level flipping ratio $r^{(k)} = 1- \overline{c}_{kk} $ and recovery distribution $\mathcal{P}_\mathcal{N}(k): P(y^j=1|\tilde{y}^k=1) = \overline{c}_{jk} / \sum_{j\neq k}\overline{c}_{jk}$, $1 \leq j \leq C$, where $r_i = r^{(k)}$ and $\mathcal{P}_\mathcal{N}(\mathbf{x}_i, \tilde{\mathbf{y}}_i)=\mathcal{P}_{\mathcal{N}}(k)$ if $\tilde{y}_i^k =1$. Taking symmetric noise as an example, it means that the label of a sample is uniformly perturbed to any other labels. In this case, symmetric noise with noise ratio $r_{sym}$ indicates that $r^{(k)}=r_{sym}$ and $\mathcal{P}_{\mathcal{N}}(k)$ is uniform distribution for all $k$. 

\label{sec:pre_lnl_cnr}
Solving LNL tasks based on the above formulation is equivalent to answering the two research questions below: which sample contains label noise and how do we correct the noise? Because the ground truth labels are in fact unknown, answering the first question implies the need to detect samples with a noisy label, that is, a good $r_i$ estimation; answering the second one further requires a good $\mathcal{P}_\mathcal{N}(\mathbf{x}_i, \tilde{\mathbf{y}}_i)$ or a simplified $\mathcal{P}_{\mathcal{N}}(k)$ estimation and is much more ambitious, difficult and highly dependent on the scale and type of label noise.

\subsection{Contrastive Learning (CL) under label noise}
\label{sec:pre_cl}
\subsubsection{Background and definition} 
Inspired by the need to detour the noise distortion in the label space, there are some attempts to leverage contrastive learning to design LNL framework \cite{yi2022learning,li2022selec}, as it has been found to be beneficial for weakly self-supervised representation learning \cite{chen2020simple, he2020momentum}. Contrastive learning aims to learn a good feature encoder network $\mathbf{z} \triangleq \phi_{\theta_{enc}}(\mathbf{x}) \in \mathbb{D}_{\mathbf{z}}$ that embeds a sample $\mathbf{x} \sim \mathcal{D}_\mathbf{x}$ closer to its augmentation $\mathbf{x}^+$ (referred to as the positive sample) than other samples (referred to as the negative samples) in the representation space $\mathbb{D}_{\mathbf{z}}$. Formally, let $g_{\theta_{proj}}:\mathbb{D}_{\mathbf{z}} \rightarrow \mathcal{S}^{N-1}$ be a projector network that maps $\mathbf{z}$ to a $N$-dimensional $L$2-normalized hypersphere space, $f(\mathbf{x}) \triangleq g_{\theta_{proj}}(\phi_{\theta_{enc}}(\mathbf{x})) \in \mathcal{S}^{N-1}$ be the projected representation of $\mathbf{x}$ and $\{\mathbf{x}_i^-\}_{i=1}^{k_1}$ be the set of negative samples, self-supervised contrastive learning optimizes the InfoNCE loss \cite{oord2018representation} with temperature $\tau$ below:
\begin{equation}
    \begin{aligned}
    \mathcal{L}_{\text{CL}} &=  -\log \frac{e^{f(\mathbf{x})\cdot f(\mathbf{x}^+)/\tau}}{e^{f(\mathbf{x})\cdot f(\mathbf{x}^+)/\tau} + \sum_{i=1}^{k_1}e^{f(\mathbf{x})\cdot f(\mathbf{x}_i^-)/\tau}}\\
    =-&\frac{f(\mathbf{x})\cdot f(\mathbf{x}^+)}{\tau} + \log (e^{\frac{f(\mathbf{x})\cdot f(\mathbf{x}^+)}{\tau}} + \sum_{i=1}^{k_1}e^{\frac{f(\mathbf{x})\cdot f(\mathbf{x}_i^-)}{\tau}}).
    \end{aligned}
    \label{eqn:pre_cl_def}
\end{equation}

Moreover, when sample labels are available, label information can also be incorporated into the contrastive learning framework \cite{khosla2020supervised} by considering the two samples of the same class as positive pairs. Let $\{\mathbf{x}_i^+\}_{i=1}^{k_1}$ be the set of positive samples that are from the same class and $\{\mathbf{x}_j^-\}_{j=1}^{k_2}$ be the set of negative samples that are from different classes, the supervised contrastive learning (SupCL) loss can be defined as:
\begin{equation}
    \begin{aligned}
    \mathcal{L}_{\text{SupCL}}=-\frac{1}{k_1}\sum_{i=1}^{k_1}\log \frac{e^{f(\mathbf{x})\cdot f(\mathbf{x}_i^+)/\tau}}{\sum\limits_{i=1}^{k_1}e^{\frac{f(\mathbf{x})\cdot f(\mathbf{x}_i^+)}{\tau}} + \sum\limits_{j=1}^{k_2}e^{\frac{f(\mathbf{x})\cdot f(\mathbf{x}_j^-)}{\tau}} }.
    \end{aligned}
    \label{eqn:pre_cl_supcl}
\end{equation}
\subsubsection{Motivation of CL for LNL}
In the second line of Eqn. (\ref{eqn:pre_cl_def}), optimizing the first term minimizes the representation distance between positive pairs while optimizing the second term maximizes the representation distance of negative pairs \cite{wang2020understanding}. 
Because the augmented view of a sample is likely to have same class label, minimizing Eqn. (\ref{eqn:pre_cl_def}) can intuitively push samples close to other samples of the same classes.
This gives the \textit{motivation} of using contrasted representations to address label noise: suppose the encoder $\phi_{\theta_{enc}}$ is perfectly learned where all samples from the same class have similar representations which are much more distinguishable than the representations of samples from other classes, a feasible LNL problem can thus be solved easily by replacing $\mathbf{x}$ with $\mathbf{z}$ in Eqn. (\ref{eqn:pre_noisechannel}) and learn $P(\mathbf{y}|\mathbf{\tilde{y}}, \mathbf{z})$ by simply fine-tuning a lightweight classifier $\mathbf{\hat{y}} \triangleq \omega_{\theta_{cls}}(\mathbf{z})$ cascaded after $\phi_{\theta_{enc}}$ with the annotated data.  

\subsubsection{Challenges} 
However, contrastive learning is limited in the LNL setting due to two key \textit{challenges}.
First, the learned encoder learned through the InfoNCE loss in Eqn. (\ref{eqn:pre_cl_def}) cannot be perfectly learned in practice with only a weakly supervised self-supervision signal. This again makes the cascaded classifier vulnerable to overfitting to the noisy labels. Second, despite the existing works \cite{khosla2020supervised, kim2020mixco} showing the furthermore potential of CL with auxiliary supervised positive pair signals such as SupCL, there remains the problem of estimating the distribution in Eqn. (\ref{eqn:pre_noisechannel}) such as inferring which sample has a reliable label as discussed in section \ref{sec:pre_lnl_cnr}. 

To this end, two main questions are on the table: (i) how to stimulate the potential of contrastive learning with the existence of noisy labels? (ii) How to fuse contrastive learning into the current classifier-based noise recovery framework efficiently and effectively? This inspires us to develop ChiMera - a unified two-stage LNL framework that efficiently combines and utilizes both the potential of contrastive learning and classifier-based noise-recovery strategy.


\section{Methods}
In this section, we present ChiMera, a unified two-stage framework (Fig. \ref{fig:chimera_flowchart}) to address the learning with noisy labels problem by leveraging contrastive representation learning and semi-supervised learning. We first provide an overview of the idea and design of ChiMera in section \ref{sec:met_overview} and discuss how to efficiently combine contrastive learning and classifier-based noise-recovery into one scheme. We then present the pretraining warm-up stage in section \ref{sec:met_stage1} and the semi-supervised noise recovery stage in section \ref{sec:met_stage2}. Through this, we also introduce MixCLR, the core adaptation we designed that can boost both the performance of contrastive pretraining and downstream noise recovery classifier. Next in section \ref{sec:met_asymix} we further discuss how we extend MixCLR to its noise-type aware variants - AsyMixCLR to further help asymmetric label noise.      

\subsection{ChiMera: A two-stage detect-then-correct framework}
\label{sec:met_overview}
Fig. \ref{fig:chimera_flowchart} shows the overview of ChiMera. ChiMera is a noise recovery framework, that is, it aims to address label noise by learning to recover $\mathbf{y}_i$ from noisy observation $(\mathbf{x}_i,\mathbf{\tilde{y}}_i) \in \mathbf{D}_{\mathcal{N}}$. ChiMera forms an intuitive two-step solution to accomplish this task: first, detect samples with noisy labels, learn to fit samples with clean labels, and extract informative representations from samples with noise labels. To achieve this, it utilizes two learning schemes, a contrastive representation learning scheme, and a semi-supervised learning-based iterative detect-then-correct scheme. To efficiently bridge the gap between these two schemes, ChiMera is designed and optimized for the following two important objectives: (i) learning a good representation against label noise and (ii) utilizing the learned representation to detect and correct label noise. ChiMera has two major learning stages, the pretraining warm-up stage, and the iterative semi-supervised noise recovery stage, because of the ability of contrastive learning to leverage self-supervision without the need to incorporate labels. However, contrastive learning is not limited to only warming up the model. Instead, in the second iterative noise recovery stage, contrastive learning is also incorporated to leverage the outcomes from the progressing noise recovery process and provide better representation learning supervision.    

We now introduce the main structure and important processes of ChiMera. The main model of ChiMera $M=\{\phi_{\theta_{enc}}, g_{\theta_{proj}}, \omega_{\theta_{cls}}\}$ has a shared backbone encoder and two heads, a projector network for contrastive learning, and a lightweight classification head for label prediction. In the first stage, the encoder $\phi_{\theta_{enc}}$ and the projector $g_{\theta_{proj}}$ are pre-trained through contrastive learning and MixCLR, a novel, simple yet effective contrastive objective performing on mixed-up samples via Mixup augmentation \cite{zhang2017mixup}. Then the classifier $\omega_{\theta_{cls}}$ is warmed up using the noisy label supervision with a small round of training. In the second stage, given $\mathbf{x}_i$ and let $\mathbf{p}_i=cls(\mathbf{x}_i)=\omega_{\theta_{cls}}(\phi_{\theta_{enc}}(\mathbf{x}_i))\in [0,1]^{C}$ be its predicted logits, ChiMera estimates $\hat{r}_i=h_{\theta_{nd}}(\mathbf{p}_i,\mathbf{\tilde{y}}_i)$, the probability that $\mathbf{\tilde{y}}_i$ is noisy using a parameterized noise detection module $h_{\theta_{nd}}$. ChiMera then divides the $\mathbf{D}_{\mathcal{N}}=\mathbf{S}_\mathcal{C} \cup \mathbf{S}_\mathcal{N}$ into two subgroups, the possibly clean subset $\mathbf{S}_\mathcal{C}$ and noisy subsets $\mathbf{S}_\mathcal{N}$. 
This simple yet crucial step explicitly concentrates cleaner label supervision and enables the possibility to apply more powerful variants of contrastive learning such as SupCL as well as semi-supervised learning. 
Moreover, if any prior, such as noise type, is known, it also brings the flexibility to exploit it to help estimate $\mathcal{P}_{\mathcal{N}}(\mathbf{x}_i, \mathbf{\tilde{y}}_i)$ or the simplified $\mathcal{P}_{\mathcal{N}}(k)$. 

As discussed before, one key advantage of ChiMera is that contrastive learning continues to help to learn representation by leveraging the outcome of the noise detection module after helping pretraining. ChiMera utilizes two major learners in stage II to refine the main model $M$, a semi-supervised noise corrector $\mathcal{L}_{NC} \leftarrow \Phi_{NC}(\mathbf{S}_\mathcal{C}, \mathbf{S}_\mathcal{N}|\phi_{\theta_{enc}}, \omega_{\theta_{cls}})$, and an enhanced contrastive learner $\mathcal{L}_{\text{CL}^+} \leftarrow \Phi_{CL^+}(\mathbf{S}_\mathcal{C}, \mathbf{S}_\mathcal{N}|\phi_{\theta_{enc}}, g_{\theta_{proj}})$. The effectiveness and efficiency of ChiMera are thus related to three key design task objectives between the two learners and across two stages: (i) good intra-learner learning, (ii) effective inter-learner fusing, and (iii) efficient optimization alignment across two stages. 
Interestingly, we find properly exploiting MixCLR can be both intuitively and practically beneficial for both stages and learners, answering these design questions in a unified way.
The core intuition of MixCLR is that it may not only improve the quality and alignment of the learned representations in both stages but also help to fuse the contrastive learner and the semi-supervised noise corrector efficiently via mixup augmentation. Moreover, we developed its variant AsyMixCLR to help better address the asymmetric label noise which usually contains more hard negatives. Next, we introduce the details of both stages and MixCLR design.

\subsection{Stage I: Pre-training and warm-up enhanced by MixCLR}
\label{sec:met_stage1}
ChiMera aims to warm up its main model $M$ from scratch in the first stage. Even though the label may be wrong, previous works have shown that warming up $M$ using noisy labels for a small number of epochs \cite{li2020dividemix, liu2020early} is beneficial to learning from the correct label subgroup. That is, let $\mathbf{p}_i=cls(\mathbf{x}_i)$ be the logits of $\mathbf{x}_i$, the encoder $\phi_{\theta_{enc}}$ and the classifier head $\omega_{\theta_{cs}}$ are warmed-up via a cross-entropy loss objective: 
 \begin{equation}
    \begin{aligned}
\mathcal{L}_{\text{warmup\_cls}} &= - \frac{1}{|\mathbf{D}_{\mathcal{N}}|}\sum^{|\mathbf{D}_{\mathcal{N}}|}_{i=1} \mathbf{\tilde{y}}_i^T\log \mathbf{p}_i.
    \end{aligned}
    \label{eqn:met_warmup_cls}
\end{equation}
However, a randomly initialized encoder $\phi_{\theta_{enc}}$ is much more sensitive to the noise compared to an encoder first pre-trained with contrastive learning. Therefore, the first stage of ChiMera is mainly about how to design an effective contrastive learning strategy for pre-training $\phi_{\theta_{enc}}$ before applying Eqn. (\ref{eqn:met_warmup_cls}). 
\subsubsection{Pre-training via MixCLR}
When considering contrastive learning, all samples $\mathbf{x} \in \mathbf{D}_{\mathcal{N}}$ will be seen as unlabeled data in order to fully avoid learning from label noise. In this situation, only weak supervision, such as augmentation-based self-supervision, can be leveraged. Let $A(.): \mathbb{D}_\mathbf{x} \rightarrow  \mathbb{D}_\mathbf{x}$ be a probabilistic augmentation function, the augmented positive samples of $\mathbf{x}$ can be generated as $\mathbf{x}^+=A(\mathbf{x})$. In practice, it is usually better to construct $\mathbf{x}^{v_1}=A(\mathbf{x})$ and $\mathbf{x}^{v_2}=A(\mathbf{x}), \mathbf{x}^{v_1} \neq \mathbf{x}^{v_2}$, two augmented views of $\mathbf{x}$ and use them as positive pairs to gain a more robust learned model.

In the ideal case that the annotations are all clean in the dataset for the downstream fine-tuning, the representation trained with standard contrastive learning objectives will already be good. 
However, under the existence of noisy labels, the resulting classification hyperplane will be more vulnerable and less robust to label noise (Fig. \ref{fig:alignment_and_tsne}a). Since mixup augmentation \cite{zhang2017mixup} can greatly improve the robustness of the model by randomly perturbing both the feature of the sample and the label by interpolating it with another pair of samples and label, we, therefore, develop MixCLR to make the learned representation more reliable for downstream label noise.

Although contrastive learning and mixup have been extensively studied in the literature \cite{verma2021towards,kim2020mixco,shen2020mix,zhang2021m,lee2020mix,zhang2020does}, the key idea of our method is to perform mixup on augmented samples, resulting in two new views to perform contrastive learning. Different from the existing methods that leverage linear interpolation alignment in the representation or projection space, MixCLR (Fig. \ref{fig:mixclr_illustration}) directly performs self-supervised contrastive learning on mixed-up augmentations from two different samples. In particular, let $\mathbf{x}_i^{v_1}, \mathbf{x}_i^{v_2}$ be two randomly sampled augmented views of $\mathbf{x}_i$, $\mathbf{x}_j^{v_1}, \mathbf{x}_j^{v_2}$ be two randomly sampled augmented views of another sample $\mathbf{x}_j$, and $\lambda \sim Beta(\alpha, \alpha)$ be the mixup ratio, where $\alpha >0$. We then create positive pairs by mixing up augmented views of $\mathbf{x}_i$ and $\mathbf{x}_j$ as:
 \begin{equation}
    \begin{aligned}
\mathbf{x}^{v_1}_{i,j}(\lambda)&= \lambda \mathbf{x}_i^{v_1} + (1-\lambda)\mathbf{x}_j^{v_1},\\ \mathbf{x}^{v_2}_{i,j}(\lambda)&= \lambda \mathbf{x}_i^{v_2} + (1-\lambda)\mathbf{x}_j^{v_2}.
    \end{aligned}
    \label{eqn:met_mixup}
\end{equation}

Here, $\mathbf{x}_i^{v_1}$($\mathbf{x}_i^{v_2}$) and $\mathbf{x}_j^{v_1}$($\mathbf{x}_j^{v_2}$) could come from different augmentation operations (e.g., $\mathbf{x}_i^{v_1}$ is rotation and $\mathbf{x}_j^{v_1}$ is cropping). MixCLR uses $\mathbf{x}^{v_1}_{i,j}$ and $\mathbf{x}^{v_2}_{i,j}$ as positive pairs in the contrastive learning framework. For brevity we use $\mathbf{x}^{v_1}_{i,j}$ and $\mathbf{x}^{v_2}_{i,j}$ to denote the mixed-up augmented views.
The negative samples can be obtained accordingly using mixed-up augmentation from another pair of images. For $\mathbf{x}^{v_1}_{i,j}$, its negative samples are defined as $\{\mathbf{x}^{v_1}_{i_l', j_l'},(i_l', j_l') \neq (i, j)\}_{l=1}^{k_3}$. We then define mix-up enhancements-based contrastive representation learning loss MixCLR as the following:
\begin{equation}
    \begin{aligned}
    \mathcal{L}_{\text{MixCLR}}&(\mathbf{x}^{v_1}_{i,j},  \mathbf{x}^{v_2}_{i,j};\lambda,\alpha) = \\ & -\log \frac{e^{f(\mathbf{x}^{v_1}_{i,j})\cdot f(\mathbf{x}^{v_2}_{i,j})/\tau}}{e^{f(\mathbf{x}^{v_1}_{i,j})\cdot f(\mathbf{x}^{v_2}_{i,j})/\tau} + \sum\limits_{l=1}^{k_3}e^{f(\mathbf{x}^{v1}_{i,j})\cdot f(\mathbf{x}^{v_1}_{i'_l,j'_l})/\tau}}.
    \end{aligned}
    \label{eqn:met_mixclr}
\end{equation}

The final pre-training loss optimizing $\theta_{enc}$ and $\theta_{proj}$ therefore contains a MixCLR loss and a vanilla contrastive learning loss as the following with hyperparameter $\lambda_{pt}$ and $\alpha$:
\begin{equation}
    \begin{aligned}
    \mathcal{L}_{\text{PT}}(\alpha) &= \mathcal{L}_{\text{CL}} + \lambda_{pt}   \mathop{\mathbb{E}}\limits_{\lambda, i, j} \mathcal{L}_{\text{MixCLR}}(\mathbf{x}^{v_1}_{i,j}, \mathbf{x}^{v_2}_{i,j}; \lambda,\alpha).
    \end{aligned}
    \label{eqn:met_PT}
\end{equation}

After warming $\phi_{\theta_{enc}}$ and $g_{\theta_{proj}}$, the classifier head $\omega_{\theta_{cls}}$ will then be trained using the loss in Eqn. (\ref{eqn:met_warmup_cls}). With a pre-trained encoder, $\omega_{\theta_{cls}}$ will be more robust to noisy labels.
\subsubsection{MixCLR improves the representation alignment and quality}
\begin{figure}[ht]
\centering
 \includegraphics[width=\linewidth]{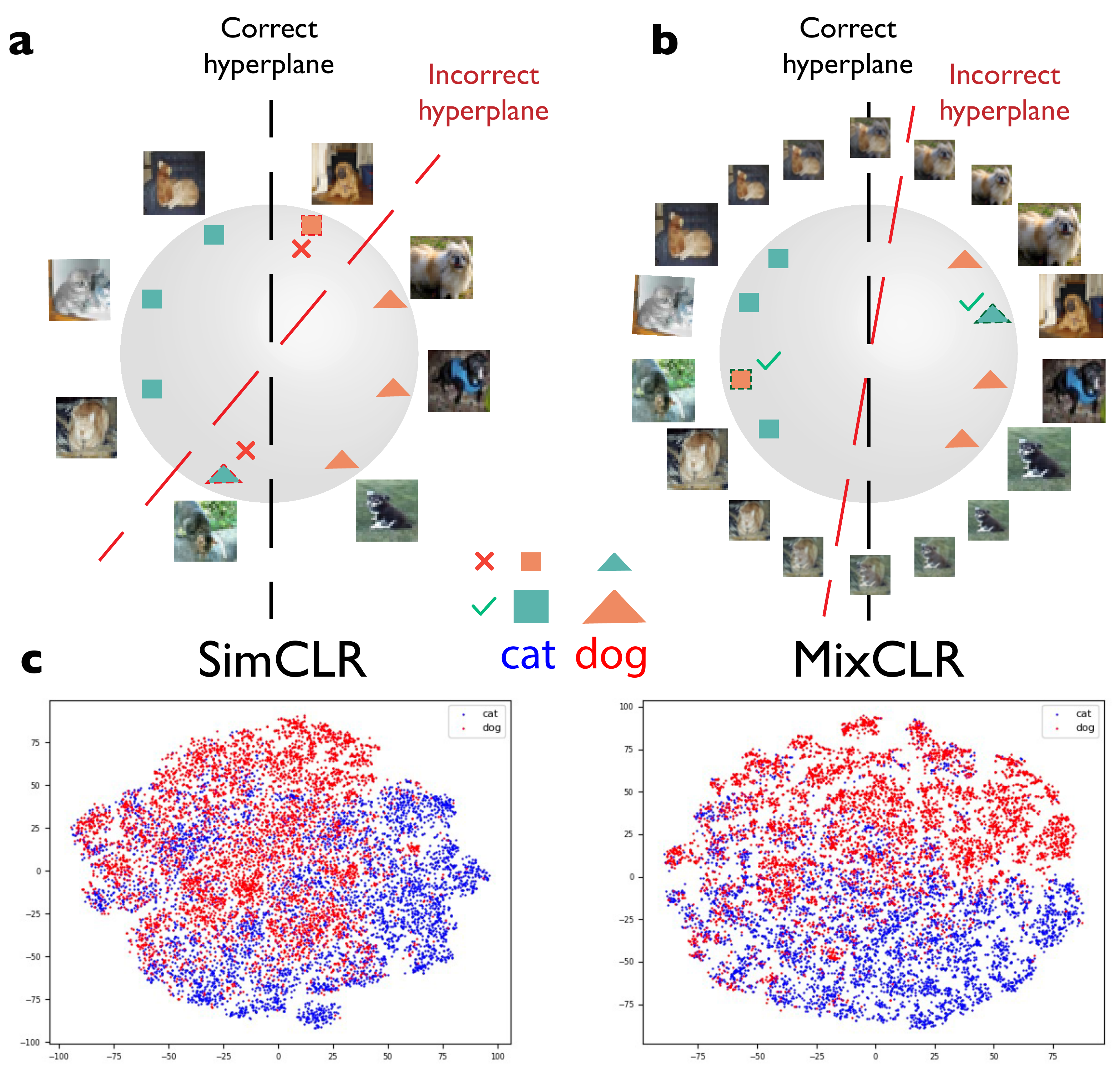}
  \setlength{\abovecaptionskip}{-0.1cm}
  \setlength{\belowcaptionskip}{-0.1cm}
\caption{\textbf{a}. Hypersphere of the vanilla contrastive learning (SimCLR). Shape denotes the true classes, and color denotes the (maybe) noisy labels. \textbf{b}. Hypersphere of MixCLR. The hypersphere is less sensitive to noisy labels because it also learns the representation of mixed image views. \textbf{c}. t-SNE visualizations of representations learned by SimCLR and MixCLR. Silhouette scores are 0.09 and 0.18.}
\label{fig:alignment_and_tsne}
\end{figure}

MixCLR loss can be regarded as applying contrastive learning to a new mixed-up dataset constructed using linear interpolation in the raw data space: $\mathbf{D}_{i,j}(\lambda):\{\mathbf{x}_{i,j}(\lambda)| \mathbf{x}_i,\mathbf{x}_j \in \mathbf{D}^{(\mathbf{x})}_{\mathcal{N}}, i\neq j\}$. We find that this key idea of directly utilizing the self-supervision signal based on mixed-up samples instead of aligning the interpolation in the latent space is crucial to its success. This is because the self-supervision signal has been shown \cite{chen2020simple,you2020graph} to be the key to the success of contrastive learning as discussed in Eqn. (\ref{eqn:pre_cl_def}), as the augmentation operation still preserves the key information about $\mathbf{x}$.
Let $\mathcal{P}_{pos}$ be the distribution of positive pairs over $\mathbb{D}_\mathbf{x} \times \mathbb{D}_\mathbf{x}$, MixCLR is able to sample more positive pairs from the place close to the intersection of different classes, i.e., the class boundaries (Fig. \ref{fig:alignment_and_tsne}b).
With the positive pairs of mixed samples, the encoder network $\phi_{\theta_{enc}}$ manages to better anchor the class boundary in the representation space, and therefore indicates a better clustering effect for the original samples. Such effect is correlated with the notion of alignment, calculated as the following with hyperparameter $\beta >0$:
\begin{equation}
    \begin{aligned}
    \mathcal{L}_{\text{alignment}}(f;\beta) = \mathop{\mathbb{E}}\limits_{(\mathbf{x}, \mathbf{x}^+)\sim \mathcal{P}_{pos}}[||f(\mathbf{x})-f(\mathbf{x}^+)||_2^{\beta}].
    \end{aligned}
    \label{eqn:met_align}
\end{equation}

\begin{figure}[ht]
\centering
 \includegraphics[width=\linewidth]{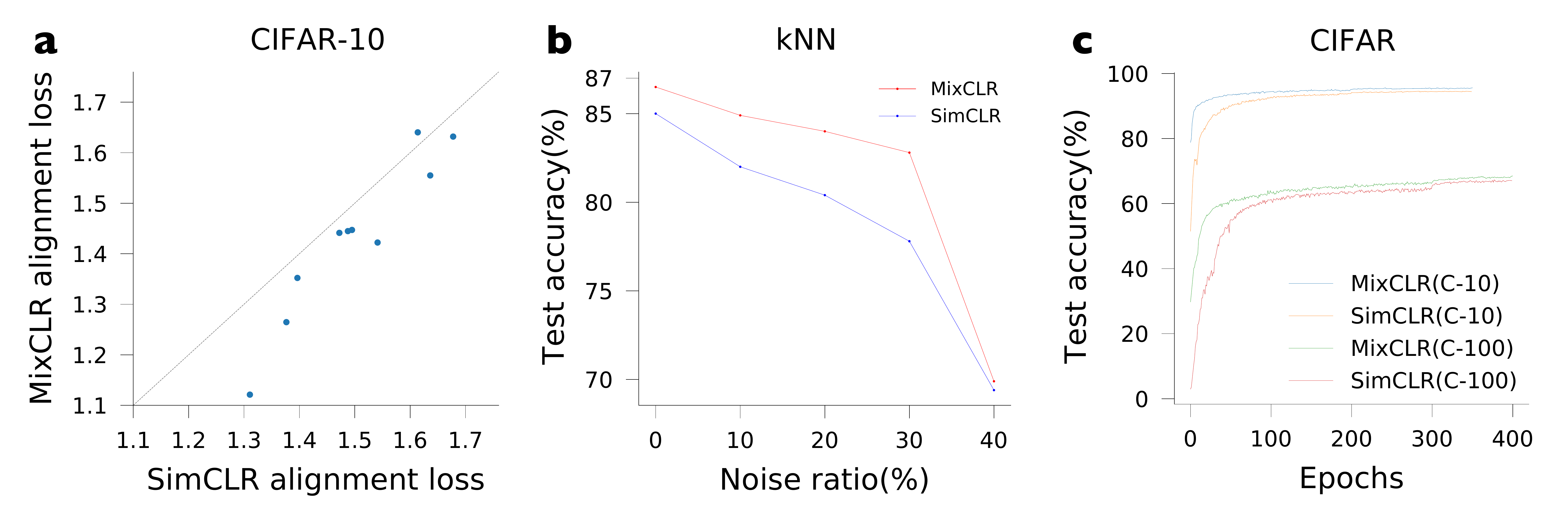}
  \setlength{\abovecaptionskip}{-0.1cm}
  \setlength{\belowcaptionskip}{-0.1cm}
\caption{\textbf{a}. Comparison of the alignment loss by SimCLR and by MixCLR on CIFAR-10. Lower alignment loss indicates better representation. Each dot is a class in CIFAR-10.  \textbf{b}. kNN performance. \textbf{c}. Comparison of the warm-up effects by MixCLR and SimCLR.}
\label{fig:align_knn_ft}
\end{figure}

Specifically, alignment states that positive pairs should have similar features, and be robust to unnecessary noisy features. We hypothesize that MixCLR improves the alignment of the original sample subgroup when noisy labels are present.
To validate our hypothesis, we calculated the intra-class alignments ($\beta=2$) for MixCLR and SimCLR on CIFAR-10 (Fig. \ref{fig:align_knn_ft}a). We found that MixCLR achieves a lower alignment loss on 9 out of 10 classes, demonstrating its effectiveness in improving the alignment. To see MixCLR learns improved representation, we obtain representations of `dog' and `cat' images by first training MixCLR / SimCLR on the whole CIFAR-10 dataset and then fine-tuning on these two categories. Fig. \ref{fig:alignment_and_tsne}) illustrates the t-SNE visualizations \cite{van2008visualizing} of the representations obtained. As one can see, MixCLR shows more uniformized clusters. Furthermore, MixCLR quantitatively obtains a higher silhouette score (0.18 VS 0.09). We also directly perform kNN using the two representations to evaluate if MixCLR representation leads to a more robust classification boundary under label noise (Fig. \ref{fig:align_knn_ft}b). MixCLR clearly outperforms SimCLR on all noise ratios. This again shows that MixCLR achieves better alignment and a more robust decision hyperplane. Therefore, the designated MixCLR representation has improved alignment and 
can learn a more unified, continuous representation manifold that can better depict the relationship between mixed-up samples and unmixed-up samples.



\subsection{Stage II: Iterative noise detection, correction, and representation refinement} 
\label{sec:met_stage2}
Given a warmed main model $M$, the second stage of ChiMera iteratively improves it with an exploitation-then-refinement pipeline. As already briefly discussed in section \ref{sec:met_overview}, the noise detector $h_{\theta_{nd}}$ utilizes the warmed classifier to divide $\mathbf{D}_{\mathcal{N}}$ into $\mathbf{S}_{\mathcal{C}}$ and $\mathbf{S}_{\mathcal{N}}$ based on the loss between the prediction and the given label. Then, a contrastive learner $\Phi_{CL^+}$ focusing on the clean subset $\mathbf{S}_{\mathcal{C}}$ and a noise corrector $\Phi_{NC}$ leveraging mixup-enhanced semi-supervised learning are used for representation refinement and noise correction, respectively. Both of them exploit mixup augmentation, and interestingly, we find that they can be connected by further exploiting MixCLR. Next, we introduce their designs in detail.

\subsubsection{Detecting label noise by validating the loss}
Following recent work \cite{li2020dividemix}, we set the noise detector $h_{\theta_{nd}}$ to be a two-component Gaussian Mixture Model (GMM) and exploit it to check the loss between the noisy labels and logits predicted by the warmed classifier after stage I. Assuming that $M$ has acquired some ability to perform the correct classification, the loss $l_i=-\mathbf{\tilde{y}}_i^T\log\mathbf{p}_i$ of sample $\mathbf{x}_i$ will be small if the label is correct and large if wrong. Therefore, the GMM model will fit two Gaussian distributions based on the loss set $\{l_i\}^{|\mathbf{D}_{\mathcal{N}}|}_{i=1}$ and use it to generate the split. 
Let $\hat{\gamma}_i \in [0,1]$ be the probability that $\mathbf{x}_i$ belongs to the distribution with a smaller mean value, i.e., the probability that $\mathbf{\tilde{y}}_i$ is clean, we have $\hat{r}_i=1-\hat{\gamma}_i$. 
To obtain a discrete split of $\mathbf{S}_{\mathcal{C}}$ and $\mathbf{S}_{\mathcal{N}}$, a threshold $\delta$ is set, and only samples with $\hat{r}_i < \delta$ will be treated as clean labels. Compared to existing detecting methods that rely on similarities in the learned representation \cite{li2022selec}, this method directly validates and exploits the ongoing ability gained by the classifier and is, therefore, more promising. 

\subsubsection{Concentrated contrastive learning with filtered label supervision}

The filtered $\mathbf{S}_{\mathcal{C}}$ provides a more reliable subset of samples with correct label supervision and therefore provides an opportunity to refine the representation quality by concentrating the contrastive learning on it. Specifically, in stage \uppercase\expandafter{\romannumeral 2}, ChiMera only performs contrastive learning on the samples from the clean subset $\mathbf{S}_{\mathcal{C}}$, including both standard contrastive learning and MixCLR. Although the MixCLR objective is still constructed on self-supervised mixed positive pairs, the standard contrastive learning objective can be replaced with the supervised contrastive learning objective defined in Eqn. (\ref{eqn:pre_cl_supcl}) to incorporate informative labels. Formally, let $\mathcal{L}^{\mathcal{C}}_{\text{SupCL}}$ and $\mathcal{L}^{\mathcal{C}}_{\text{MixCLR}}(\lambda, \alpha)$ be the supervised contrastive loss and the MixCLR loss concentrated on $\mathbf{S}_{\mathcal{C}}$, the adapted loss objective $\mathcal{L}_{\text{CL}^+}$ with the hyperparameter $\lambda_{cl}$ and $\lambda_{mix}$ is the following: 
\begin{equation}
    \begin{aligned}
    \mathcal{L}_{\text{CL}^+}(\lambda_{cl}, \lambda_{mix}) &= \lambda_{cl}\mathcal{L}^{\mathcal{C}}_{\text{SupCL}} + \lambda_{mix}   \mathcal{L}^{\mathcal{C}}_{\text{MixCLR}}(\lambda,\alpha).
    \end{aligned}
    \label{eqn:met_cl_refine}
\end{equation}
The only exception is if the prior that the noise type is asymmetric is known, we will stick to self-supervised contrastive objectives and add another AsyMixCLR objective to help, which we will cover in section \ref{sec:met_asymix}.

This simple adaptation is efficient and effective. First, the main adaptations are selecting samples with more informative label supervision and not relying on other complex design changes on contrastive objectives or introducing extra computational burdens. Second, concentrating on the possibly clean sample subset enables the model to learn from supervised contrastive learning or other label-guided designs. As in the learning with noisy label setting, we care more about the representations that are helpful to the classification, incorporating label supervision is a non-trivial progress. Although self-supervision positive signals leveraged in stage I might also introduce class-irrelevant information such as low-level texture information \cite{chen2020simple}, filtered label information can concentrate the representation to its class centroid. Third, applying MixCLR on $\mathbf{S}_{\mathcal{C}}$ will also adapt the representation to focus more on mixed-up samples with clear label information. Even if MixCLR does not directly use the interpolated labels, this will also be helpful for the semi-supervised noise corrector.

\subsubsection{Semi-supervised noise correction via label diffusion}
The noise corrector $\Phi_{NC}$ tries to diffuse label information from samples with clean labels to samples with noisy labels by assigning a mixed-up label to mixed-up enhancements as well as pursuing consistency between unlabeled samples and mixed-up augmentations, inspired by the recent success achieved in semi-supervised learning \cite{berthelot2019mixmatch}.
Specifically, let $(\mathbf{x}_i, \mathbf{\tilde{y}}_i) \sim \mathbf{S}_{\mathcal{C}}$ be a possibly clean sample and the label pair sampled from $\mathbf{S}_{\mathcal{C}}$, $\mathbf{u}_j \sim \mathbf{S}_{\mathcal{N}}$ be a possibly noisy sample sampled from $\mathbf{S}_{\mathcal{N}}$. Let $\mathbf{p}_j= cls(\mathbf{u}_j)$ be the logits of sample $\mathbf{u}_j$ predicted by the classifier, and let $\mathbf{x}_{i,j}(\lambda)=\lambda \mathbf{x}_i + (1-\lambda)\textbf{u}_j$ and $\mathbf{p}_{i,j}(\lambda)= cls(\mathbf{x}_{i,j}(\lambda))$ be the mixed-up samples between $\mathbf{x}_i$ and $\textbf{u}_j$ with the one-sided mixup factor $\lambda =\max(\lambda_1,1-\lambda_1), \lambda_1 \sim Beta(\alpha, \alpha)$ and the predicted logits of $\mathbf{x}_{i,j}(\lambda)$, the label diffusion objective is formulated as:
\begin{equation}
    \begin{aligned}
    \mathcal{L}_{\rm Diff}(\alpha) = \mathop{\mathbb{E}}\limits_{ \lambda, i,  j }&[-(\lambda \mathbf{\tilde{y}}_i +(1-\lambda) \textbf{p}_j)\log \textbf{p}_{i,j}(\lambda)  \\
    + &\lambda_u((1-\lambda)  \mathbf{\tilde{y}}_i+  \lambda \textbf{p}_j - \textbf{p}_{i,j}(1-\lambda) )^2 ].
    \end{aligned}
    \label{eqn:met_labeldif}
\end{equation}



\begin{algorithm}
\renewcommand{\algorithmicrequire}{\textbf{Input:}}
\renewcommand{\algorithmicensure}{\textbf{Output:}}
\renewcommand{\algorithmiccomment}{ \ \ \ // }
\caption{ChiMera Algorithm}
\begin{algorithmic}[1]
\REQUIRE Dataset $\mathbf{D}_\mathcal{N}$, pre-training, warm-up and stage \uppercase\expandafter{\romannumeral 2} epochs $E_{pre}, E_w, E_{semi}$ and batch size $B_1, B_2$, \textit{mode} =$\{normal, asym\}$,
 temperature $\tau$, sampling hyperparameter $\alpha$, noise correction iteration number ITER.
 \ENSURE Optimized model $M$.
\STATE Randomly initialize $M=\{\phi_{\theta_{enc}}, g_{\theta_{proj}}, \omega_{\theta_{cls}}\}$.
\FOR{$e=0$ to $E_{pre}$} 
    \STATE Draw random raw data batch $\{\mathbf{x}_i\}^{2B_1}_{i=1}$ from $\mathbf{D}_x$. 
    \STATE Construct augmented views $\{(\mathbf{x}^{v_1}_i,\mathbf{x}^{v_2}_i)\}$ for each  $\mathbf{x}_i$.
    \STATE Sample $\{\lambda_i\}_{i=1}^{B_1} \overset{\mathrm{iid}}{\sim} \text{Beta(}\alpha,\alpha)$ and construct mixup batch $\{(\tilde{\mathbf{x}}^{v_1}_i, \tilde{\mathbf{x}}^{v_2}_i)\triangleq (\mathbf{x}^{v_1}_{2i-1,2i}(\lambda_i),\mathbf{x}^{v_2}_{2i-1,2i}(\lambda_i))\}_{i=1}^{B_1}\}$. 

    \STATE Optimize $\phi_{\theta_{enc}}, g_{\theta_{proj}}$ with Eqn. (\ref{eqn:met_PT}). 
\ENDFOR

\STATE Warm up $\omega_{\theta_{cls}}$ with Eqn. (\ref{eqn:met_warmup_cls}) for $E_w$ epochs.
\FOR{$e =0 $ to $E_{semi}$}
    \STATE Get  $\mathbf{S}_\mathcal{C}, \mathbf{S}_\mathcal{N}=$GMM$(\mathbf{D}_\mathcal{N}, M)$. 
    \FOR{iter = 0 to ITER}
    \STATE Draw data batch $\hat{X}_B\triangleq\{\hat{\mathbf{x}}_i, \hat{\mathbf{y}}_i\}_{i=1}^{2B_2}$ from $\mathbf{S}_\mathcal{C}$.
    \STATE Draw data batch $\hat{U}_B\triangleq\{\hat{\textbf{u}}_{i}\}_{i=1}^{2B_2}$ from $\mathbf{S}_\mathcal{N}$.  
    \STATE Compute $\mathcal{L}_{\text{MixCLR}}$ on $\{\hat{\mathbf{x}_i}\}^{2B_2}_{i=1}$ with Eqn. (\ref{eqn:met_mixclr}).


    \STATE \textbf{if} \textit{mode} is \textit{normal} \textbf{do}
    \STATE \ \ Compute $\mathcal{L}_{\text{CL}^+}$ on $\{\hat{\mathbf{x}_i}\}^{2B_2}_{i=1}$ with Eqn. (\ref{eqn:met_cl_refine}).
    \STATE \textbf{else if} \textit{mode} is \textit{asym} \textbf{do} // See algorithm \ref{alg:Infer}.
    \STATE \ \ Infer Subset Partition $\Pi_S = \text{Infer}(\mathbf{D}_\mathcal{N}, M)$.
    \STATE \ \ Construct hard sample pairs batch $\{\hat{\mathbf{h}_i}\}^{2B_2}_{i=1}$ based on $\Pi_S$ and compute AsyMixCLR loss $\mathcal{L}_{\text{AsyMix}}$. 
    \STATE \ \ Compute $\mathcal{L}_{\text{CL}^+}$ on $\{\hat{\mathbf{x}_i}\}^{2B_2}_{i=1}$ with Eqn. (\ref{eqn:met_cl_asymix}).
    \STATE \textbf{end if}
    \STATE Calculate $\mathcal{L}_{NC}$ using $\hat{X}_B$ and $\hat{U}_B$ with Eqn. (\ref{eqn:met_noise_corrector}).
    \STATE Optimize $M$ with $\mathcal{L}= \mathcal{L}_{NC} + \mathcal{L}_{\text{CL}^+}$. 
    \ENDFOR
\ENDFOR
\end{algorithmic}
\label{alg:ALG1}
\end{algorithm}

By optimizing the classifier in $M$ using Eqn. (\ref{eqn:met_labeldif}), it is encouraged to assign the interpolated label to the mixed-up samples. This is the key learning objective that enables ChiMera to perform semi-supervised noise correction. 
Suppose $\mathbf{x}_i$ and $\mathbf{u}_j$ are of the same class, i.e., $\mathbf{y}_i =\mathbf{y}_j$, the mixed samples $\mathbf{x}_{i,j}(\lambda)$ and $\mathbf{x}_{i,j}(1-\lambda)$ should still have the optimization target $\mathbf{y}_{i,j}(\lambda)=\mathbf{y}_{i,j}(1-\lambda)=\mathbf{y}_i=\mathbf{y}_j$. In this case, the optimal choice of $\mathbf{p}_j$, $\mathbf{p}_{i,j}(\lambda)$, and $\mathbf{p}_{i,j}(1-\lambda)$ will all be $\mathbf{\tilde{y}}_i$. Suppose $\mathbf{\tilde{y}}_i$ is clean, it is equivalent to performing a perfect classification on $\mathbf{u}_j$. When $\mathbf{y}_i \neq \mathbf{y}_j$, an easy way to achieve optimal for Eqn. (\ref{eqn:met_labeldif}) is to always predicts $\mathbf{p}_{i,j}(\lambda)=\lambda \mathbf{\tilde{y}}_i + (1-\lambda)\mathbf{p}_j$ and vice versa for $\mathbf{p}_{i,j}(1-\lambda)$. Therefore, minimizing the label diffusion loss in both cases will lead to perfect classification performance. This indicates why MixCLR can boost label diffusion, as it helps to refine the representation of the mixed-up augmentations. Moreover, in practice, $\mathbf{x}_i$ and $\mathbf{u}_j$ will be randomly augmented to $\mathbf{\tilde{x}}_i=A(\mathbf{x}_i)$ and $\mathbf{\tilde{u}}_j=A(\mathbf{u}_j)$ before applying the mixup to achieve better label diffusion performance. Applying MixCLR thus naturally fits this situation, as its pair construction process is similar.

To clearly see the superiority of performing MixCLR during stage \uppercase\expandafter{\romannumeral 2}, we assessed the symmetric label noise on complete CIFAR-10 and CIFAR-100 (Fig. \ref{fig:align_knn_ft}c) to check if performing MixCLR can make the classification easier under noisy labels. We found that MixCLR achieved much higher test accuracy in the first few epochs as well as higher final accuracy compared to the ablation experiment that only performs SimCLR consistently, indicating that MixCLR accelerates the label diffusion process even when noisy labels are present.

In reality, when the number of classes $C$ is large, it is more possible to encounter the second case when selecting $\mathbf{x}_i$ and $\mathbf{u}_j$, as even for a balanced training dataset after the split, the probability of having $\mathbf{y}_i=\mathbf{y}_j$ is $1/C$. Therefore, effective label diffusion requires the success of predicting interpolated mixed-up augmentations, that is, $\textbf{p}_{i,j}(\lambda)$ and $\textbf{p}_{i,j}(1-\lambda)$. This further suggests the superiority to allow MixCLR to focus on those mixed-up enhancements from $\mathbf{S}_{\mathcal{C}}$, i.e., samples with clear label information. The refined representation can better anchor on the filtered label information and thus resist the label noise better, relieving $M$ from overfitting to the degraded representation of $\mathbf{u}_j$ which leads to arbitrary predicted logits $\mathbf{p}_j$ and $\mathbf{p}_{i,j}(\lambda)$ by simply satisfying $\mathbf{p}_{i,j}(\lambda)=\lambda \mathbf{\tilde{y}}_i + (1-\lambda)\mathbf{p}_j$.    

In addition to label diffusion loss, $\Phi_{NC}$ incorporates another group-level regularization loss following \cite{li2020dividemix} to encourage the average prediction from the noise corrector to fit the dataset distribution prior $\bm{\pi}=[\pi_1, ..., \pi_C]^T \geq 0, \sum_{c=1}^{C}\pi_c=1$ if it is known. 
The final objective for $\Phi_{NC}$ is thus the following:
\begin{equation}
    \begin{aligned}
    \mathcal{L}_{NC}=\mathcal{L}_{\rm Diff}(\alpha) +\lambda_{reg}  \bm{\pi}^T\log(|\mathbf{D}_{\mathcal{N}}| / \sum\nolimits_{ i=1}^{|\mathbf{D}_{\mathcal{N}}|}\mathbf{p}_i) .
    \end{aligned}
    \label{eqn:met_noise_corrector}
\end{equation}

Combining the objectives for $\Phi_{CL^+}$ and $\Phi_{NC}$ together, we will get the final objectives for ChiMera. Because of the computational burden, the practical loss computation is applied on mini-batches. We thus provide a 
detailed ChiMera algorithm illustration in Alg. (\ref{alg:ALG1}) to explicit show the pipeline. 

\begin{table*}[ht]
\setlength{\abovecaptionskip}{-0.5mm}
\setlength{\belowcaptionskip}{-1mm}
\begin{center}
\caption{Comparison on CIFAR-10 and CIFAR-100 by simulating symmetric noise with different noise ratios.  }
\label{table:cifar-sym}
\begin{tabular}{l|lllll|lllll}
\hline\noalign{\smallskip}
Methods (Acc(\%))&\multicolumn{5}{c|}{CIFAR-10} & \multicolumn{5}{c}{CIFAR-100}\\
\hline
Noise Ratio & 20\% & 50\%  & 80\% & 90\% & Avg. & 20\% & 50\% & 80\% & 90\% & Avg.\\
\hline

Cross-Entropy & 86.84 & 79.41 & 62.92 & 42.74 & 67.98 & 62.02 & 46.73 & 19.89 & 10.11 & 34.69\\
$\text{ELR}_+$(ResNet-34) & 95.83 & 94.88 & 93.31 & 78.73 & 90.69 & 77.61 & 73.60 & 60.82  & 33.41 & 61.36\\
Semi(PES) \cite{bai2021understanding} & 95.89 & 94.52 & 93.14 & 71.47 & 88.76 & 77.43 & 74.31 & 61.57 & 39.47 & 63.20 \\
DivideMix \cite{li2020dividemix} & 96.13 & 94.63 & 93.24  & 76.01 & 90.00 & 77.28 & 74.58 & 60.26  & 31.53 & 60.91 \\
ProtoMix \cite{li2021learning} & 95.81 & 94.31 & 92.40 & 74.98 & 89.38 & 79.07 & 74.84 & 57.75 & 29.28 & 60.24 \\
NGC \cite{wu2021ngc}& 95.88 & 94.54 & 91.59 & 80.46 & 90.62 & 79.31 & 75.91 & 62.70 & 29.76 & 61.92 \\
Sel-CL+ \cite{li2022selec} & 95.48 & 93.91 & 89.25 & 81.87 & 90.13 & 76.54 & 72.42 & 59.59 & 48.81 & 64.34\\
LRR \cite{li2021learning} & 95.85 & 94.57 & 92.39 & 90.17 & 93.25 & 79.43 & 75.01 & 65.47 & 51.32 & 67.81 \\
OT-Filter~\cite{feng2023ot} & 96.01 & 95.33 & 94.03 & 90.57 & 93.99 & 76.72 & 73.89 & 61.87 & 42.87 & 63.84\\
DM-AugDesc \cite{nishi2021augmentation} & 96.31 & 95.39 & 93.77  & 91.89 & 94.34 & 79.48 & 77.23 & 66.43  & 41.27 & 66.10\\
REED \cite{zhang2020decoupling} & 95.78 & 95.42 & 94.36 & 93.57 & 94.78 & 77.03 & 72.78 & 65.64 & 54.31 & 67.44 \\


\hline

\textbf{ChiMera (Ours)} & \textbf{96.99} & \textbf{96.57} & \textbf{95.62} & \textbf{93.71 }& \textbf{95.72} & \textbf{81.44} & \textbf{79.62} & \textbf{68.49}& \textbf{56.28} & \textbf{71.46}\\
\hline
\end{tabular}
\end{center}
\vspace{-2mm}
\end{table*}

\subsection{AsyMixCLR: Applying MixCLR to address asymmetric label noise}
\label{sec:met_asymix}


Asymmetric noise is a hard type of label noise. It only perturbs labels within a predefined subset of similar classes (e.g., cat image can only be assigned a dog label, but never a car.) and usually results in hard negatives.
Most of the existing learning with noisy label approaches do not explicitly model the asymmetric noise \cite{liu2020early,bai2021me,li2021learning}. However, real-world applications often present substantial amount of asymmetric noise. We therefore seek to address asymmetric noise by inferring the similar classes subsets and presenting AsyMixCLR to contrast the hard negatives in each subsets. 

We first formally define the asymmetric noise. Let $\Pi_S$ be a predefined partition on label classes, which divide all classes into $T$ non-overlapped subset $S_{1,\dots,T} \subset [C]$, where $S_i \cap S_j = \varnothing, i \neq j$ and $S_1 \cup S_2 \cup \dots \cup S_T = [C]$. Asymmetric label noise indicates that all the label perturbation will only occur in each subset $S_t$, $1\leq t\leq T$. Specifically, given a triplet $(\mathbf{x}_i,\mathbf{y}_i, \tilde{\mathbf{y}}_i)$, where $\mathbf{y}_i$ is the correct label of $\mathbf{x}_i$ and $\tilde{\mathbf{y}}_i$ is the perturbed label under asymmetric noise. If $\mathbf{y}_i \in S_t$, then $\mathbf{\tilde{y}}_i \in S_t $. 
\begin{algorithm}
\renewcommand{\algorithmicrequire}{\textbf{Input:}}
\renewcommand{\algorithmicensure}{\textbf{Output:}}
\renewcommand{\algorithmiccomment}{ \ \ \ // }
\caption{Inferring Subset Partitions}
\begin{algorithmic}[1]
\REQUIRE  Dataset with logits $\{(\mathbf{x}_{i},\mathbf{p}_{i})\}_{i=1}^{|\mathbf{D}_{\mathcal{N}}|}$, class number $C$, classifier $\omega_{\theta_{cls}}$, threshold $\tau_{AsyM}$, subset maximum number K, priority queue Q, initial subset partition $\Pi_S = \varnothing$.
\ENSURE Final subset Partition $\Pi_S$.
\FOR{$i=1,\dots,|\mathbf{D}_{\mathcal{N}}|$}
\STATE Get sorted list $\hat{\mathbf{p}}_i=[p_i^{j_i(1)},\dots,p_i^{j_i(\mathbf{C})}]$ 
\FOR{$k=0, \dots, K$}
    \STATE \textbf{if} $\sum_{c=1}^{k}p_i^{j_i(c)} \geq \tau_{AsyM} $
    
    \STATE \ \ \ \  Update the occurrence of $\{j_i(1),\dots,j_i(k)\}$ in Q.
\ENDFOR
\ENDFOR
\WHILE{Q is not empty}
\STATE Pop partition with top frequency $S'$ from Q.
\STATE ADD = True
\FOR{$S \in \Pi_S$}
\STATE \textbf{if} $S' \cap S \neq \varnothing$:  \ ADD = False

\ENDFOR
\STATE \textbf{if} ADD = True:
\STATE \ \ \ \ Add $S'$ into $\Pi_S$.
\ENDWHILE
\STATE Compose $S_r$ as the unseen classes, and add $S_r$ into $\Pi_S$.  
\end{algorithmic}
\label{alg:Infer}
\end{algorithm}

Suppose $\Pi_S$ is known, to incorporate this prior into the ChiMera framework, we propose AsyMixCLR, an adaptation of MixCLR which only mixes up augmentations of samples whose classes are from the same subset $S_t$. Since this loss uses the label information, we only include it in stage II and only apply it to the clean label set following the design. Specifically, let $(\mathbf{x}_i,\mathbf{\tilde{y}}_i)$ be a sample in the possibly clean label set and $\mathbf{\tilde{y}}_i \in S_t$. Then the augmentation of $\mathbf{x}$ is only mixed up with the augmentation of $\mathbf{x}_j$, where $\mathbf{\tilde{y}}_j \in S_t$. Given $\Pi_S$ and a minibatch $\{\mathbf{x}_i, \tilde{\mathbf{y}}_i\}_{i=1}^{2B}$, we first cluster the samples by which subset its noisy label belongs to. By doing this, we will get $|T|$ clusters. We then randomly pair the samples that belong to the same clusters (we re-use one sample if the number of samples in the cluster is odd). After that, we get $B$ hard pairs and apply MixCLR loss (Eqn. (\ref{eqn:met_mixclr})) on these pairs. Moreover, since asymmetric label noise has more hard negatives in each subset, indicating the lower performance of the noise detector $h_{\theta_{nd}}$, we therefore stick to the self-supervised contrastive objective $\mathcal{L}^{\mathcal{C}}_{\text{CL}}$ focusing on $\mathbf{S}_{\mathcal{C}}$. 
Let $\mathcal{L}^{\Pi_S}_{\text{AsyMixCLR}}$ be the AsyMixCLR loss applied on $\Pi_S$, the adapted objective $\mathcal{L}_{CL^+}$ is thus defined as follows:
\begin{equation}
    \begin{aligned}
    \mathcal{L}_{\text{CL}^+} = \lambda_{cl}\mathcal{L}^{\mathcal{C}}_{\text{CL}} + \lambda_{mix}   \mathcal{L}^{\mathcal{C}}_{\text{MixCLR}} + \lambda_{asym}   \mathcal{L}^{\Pi_S}_{\text{AsyMixCLR}}.
    \end{aligned}
    \label{eqn:met_cl_asymix}
\end{equation}

AsyMixCLR is more robust to asymmetric noise because it focuses on refining the representations between hard negatives in each subset. However, the prior knowledge of $\Pi_S$ is not often available in real-world applications. We therefore develop an algorithm to infer it directly from the data. To acquire such a partition, we first conduct a group-level analysis. Given $\mathbf{p}_i$, the logit prediction of a sample $\mathbf{x}_i$ output by the model $M$, we first get the sorted list $\mathbf{p}^{j_i(c)}_i$ where $\mathbf{p}^{j_i(1)}_{i} \geq \mathbf{p}^{j_i(2)}_{i} \dots \geq \mathbf{p}^{j_i(\mathbf{C})}_{i}$. We then iteratively check if $\sum_{c=1}^{k} p^{j_i(c)}_i \geq \tau_{AsyM}$ holds for $k=1,\dots,K$, where $\tau_{AsyM} \in (0,1)$ is a hard-coded threshold. If the threshold is hit, the occurrence of valid subset $\{j_i(1),\dots, j_i(k)\}$ will be recorded in a frequency-based priority queue $Q$. In practice, we find $\tau_{AsyM}=0.9$ works for most applications. We then use a greedy approach to extract class subsets from Q as partitions, restricting them to be non-overlapped. We also compose the rest of the unseen classes (if there are any) as a new subset and add it to $\Pi_S$ to make sure the partition is complete. More details can be found in (Algorithm \ref{alg:Infer}).

\section{Experimental setting}
\subsection{Simulating different types of label noises}
Though detecting and inferring the noise type is difficult in real-world settings, understanding and learning with them is helpful to evaluate the effectiveness of ChiMera by providing controllable simulation. We simulate three common types of label noise, symmetric, asymmetric, and instance-dependent noise (IDN) through the inverse of Eqn. (\ref{eqn:pre_noisechannel}). As already discussed in section \ref{sec:pre_lnl_nt} and \ref{sec:met_asymix}, a sample will be assigned a noisy label with flipping ratio $r$, symmetric noise randomly assigning a label from all classes with equal probability and asymmetric noise only randomly assigning a label from similar classes (i.e., partitioned subgroups). While these two types of noise assume the same flipping ratio over all samples, instance-dependent noise assumes instances to have different noise flipping ratios depending on their specific features \cite{xia2020part}. We evaluated ChiMera on these three types of label noises (both simulated and real) to thoroughly understand its effectiveness.

\setlength{\tabcolsep}{1.0pt}
\begin{table}[ht]
\setlength{\abovecaptionskip}{-0.5mm}
\setlength{\belowcaptionskip}{-1mm}
\begin{center}
\caption{Comparison on CIFAR-10 and CIFAR-100 by simulating instance dependent noise (IDN) with different noise ratios. }
\label{table:cifar-instance}
\begin{tabular}{l|cc|cc}
\hline\noalign{\smallskip}
\textbf{Methods (Acc (\%))}&\multicolumn{2}{c|}{\textbf{CIFAR-10}} & \multicolumn{2}{c}{\textbf{CIFAR-100}}\\

\hline
\noalign{\smallskip}
Noise Ratio & IDN-20\%   & IDN-40\%  & IDN-20\%  & IDN-40\%\\
\hline\noalign{\smallskip}

Cross-Entropy  & 87.5 ± 0.5  & 78.9 ± 0.7  & 56.8 ± 0.4  & 48.2 ± 0.5 \\
MixUp \cite{zhang2017mixup} & 93.3 ± 0.2  & 87.6 ± 0.5  & 67.1 ± 0.1  & 55.0 ± 0.1\\
DivideMix \cite{li2020dividemix}  & 95.5 ± 0.1  & 94.5 ± 0.2  & 75.2 ± 0.2 & 70.9 ± 0.1 \\
$\text{ELR}_+$(ResNet-34) \cite{liu2020early}  & 94.9 ± 0.1  & 94.3 ± 0.1 & 75.8 ± 0.2  & 74.3 ± 0.3 \\

Semi(PES) \cite{bai2021understanding} & 95.9 ± 0.1 & 95.3 ± 0.1 & 77.6 ± 0.3  & 76.1 ± 0.4 \\
TSCSI\_IDN \cite{zhao2022centrality} & 93.7 ± 0.1 & 95.0 ± 0.1 &  79.6 ± 0.2 &76.6 ± 0.2\\
DISC \cite{li2023disc}  &96.5 ± 0.1 &96.0 ± 0.1  &80.1 ± 0.1 &78.4 ± 0.2\\
\hline
\textbf{ChiMera (Ours)}  & \textbf{96.9 ± 0.1} &\textbf{96.5 ± 0.2} &  \textbf{81.5 ± 0.1} &\textbf{ 79.7 ± 0.2}\\
\hline
\end{tabular}
\end{center}
\vspace{-4mm}
\end{table}

\subsection{Datasets}
\label{sec:dataset}
We used the following datasets for evaluation. 
\textbf{CIFAR-10} and \textbf{CIFAR-100}\cite{krizhevsky2009learning} are two clean label datasets with an image size of $32\times32$, which have been widely used to simulate learning with noisy labels \cite{li2020dividemix,liu2020early,bai2021me}. Following these previous works, we used all 50k training images and 10k test images. We simulated both symmetric, asymmetric, and instance-dependent label noise with different noise ratios to test the performance and robustness of our methods. For asymmetric noise, we partition 10 classes in CIFAR-10 into five 2-component subgroups following \cite{li2020dividemix}. For instance-dependent noise, we follow the \textsl{part-dependent} label noise assumption proposed in \cite{xia2020part} which assumes that the noise of an instance depends only on its parts to avoid learning the $\textsl{ill-posed}$ transition matrix by only exploiting noise data \cite{berthon2021confidence}. 
We then apply the same strategy \cite{bai2021understanding,xia2020part} to generate instance-dependent label noise on CIFAR-10 and CIFAR-100 with different noise ratios. 
\textbf{CIFAR-10N} and \textbf{CIFAR-100N} \cite{wei2021learning} are curated noisy label datasets. They contain the same training and testing images as CIFAR-10 and CIFAR-100. Training images in CIFAR-10N span five different real-world noisy labels settings, including three noisy labels obtained from human annotators via Internet (R1, R2, R3, noise ratio 17.23\%, 18.13\%, 17.64\%), one aggregated noisy label via majority voting using three human annotators (Agg., lowest noise ratio 9.03\%), and another aggregated noisy label where the label is randomly selected from a wrong label if any of the three random labels is wrong (Worst, highest noise ratio, 40.21\%). Training images in CIFAR-100N contain a noisy label obtained from human annotators with an overall noise ratio of 40.20\%. We use it to test different real-world noisy label settings. 
\textbf{Clothing1M} \cite{xiao2015learning} and \textbf{mini-WebVision} \cite{li2017webvision} are two large-scale noisy label datasets. Clothing1M contains 1M noisy training images crawled from shopping websites and annotated according to the related textual description, and 15K clean validation images, and 10K clean test images. The estimated noise ratio in training images is 38.46\%. mini-WebVision contains $2.4M$ images with the size of $299\times299$ after resizing, collected by searching within the 1,000 concepts in ImageNet ILSVRC12 on the Internet. The estimated noise ratio is 20\%. 
\textbf{ANIMAL-10N} \cite{song2019selfie} is a noisy label dataset that contains 50K training images with noisy labels and 5K test images with clean labels. There are 5 pairs of confusing animal classes. Most of the noise occurs between the confusing pairs. The noisy ratio is 8\%. We use it to test how ChiMera can address asymmetric noise with the novel AsyMixCLR loss. 

\begin{table}[t]
    \setlength{\abovecaptionskip}{-0.5mm}
    \setlength{\belowcaptionskip}{-1mm}
    \centering
    \caption{Comparison between existing image mixture-based contrastive learning methods and MixCLR under different noise types and ratios on CIFAR-10, including the average performance on all ratios. `S/' or `A/' denotes symmetric or asymmetric noise.}

    \begin{tabular}{l|ccccc|c}
    \hline\noalign{\smallskip}
        Noise ratio & S/20\% & S/50\% & S/80\% & S/90\% & A/40\% & Average \\
        \hline\noalign{\smallskip}
        Mix-Co~\cite{kim2020mixco} & 96.18 & 95.31 & 93.76 & 90.72 & 93.88 & 93.97 \\
        M-Mix~\cite{zhang2022m} & 96.22 & 95.89 & 94.77 & 91.35 & 94.05 & 94.46 \\
        Un-Mix~\cite{shen2020mix} & 96.54 & 96.21 & 95.05 & 91.23 & 94.51 & 94.71 \\
        i-Mix~\cite{lee2020mix} & 96.86 & 96.34 & 94.89 & 91.79 & 94.36 & 94.85 \\
        \textbf{MixCLR} & \textbf{96.99} & \textbf{96.57} & \textbf{95.62} & \textbf{93.71} & \textbf{95.61} & \textbf{95.70} \\
    \hline
    \end{tabular}
    \label{tab:i-mix}
\end{table}

\subsection{Comparison approaches and implementation details}
\label{sec:implement}
We consider the following LNL methods as comparison approaches that do not require the supervision of the clean validation set: DivideMix \cite{li2020dividemix}, ELR$_+$ \cite{liu2020early}, DM-AugDesc \cite{nishi2021augmentation}, LRR \cite{li2021learning}, REED \cite{zhang2020decoupling}, Me-Momentum \cite{bai2021me}, PES(Semi) \cite{bai2021understanding}, CNLCU-H \cite{xia2021sample}, Cores \cite{cheng2020learning}, CAL \cite{Zhu_2021_CVPR}, LongReMix \cite{cordeiro2021longremix}, GJS \cite{englesson2021generalized}, C2D \cite{zheltonozhskii2021contrast}, and OT-Filter~\cite{feng2023ot}. We also compare with standard cross-entropy loss and mix-up augmentation for ablation studies. We used the released code of these methods and followed their hyperparameter recommendation. For all benchmarks, the top-1 classification accuracy(\%) is reported.

When implementing ChiMera, existing tricks for semi-supervised noise correction can be easily plugged in as its design is flexible. Following DivideMix \cite{li2020dividemix} and MixMatch \cite{berthelot2019mixmatch}, our full ChiMera model contains two instances random initialized with different seeds to perform enhancing operations such as co-divide, co-refinement, co-guessing, and sharpening. When conducting experiments on asymmetric noise and instance-dependent noise, a negative entropy term was added to introduce a confidential penalty and alleviate the model from quickly over-fitting and generating over-confident predictions.      

We followed the implementation of SimCLR \cite{chen2020simple} to realize contrastive learning. We set $\tau$ to 0.5 for all contrastive learning objectives. For all the benchmarks, we set the batch size to 1024 during the contrastive learning pretraining stage. To implement MixCLR loss, for a mini-batch of 2$m$ samples $\{\mathbf{x}_i\}^{2m}_{i=1}$, we randomly paired the samples in the mini-batch into $m$ pairs. We then mixed them up to obtain a mixed batch with size $m$ by sampling $m$ different $\{\lambda_i\}^{m}_{i=1}$. We then applied MixCLR to this new batch. 
For pre-training on CIFAR-10(N) and CIFAR-100(N), we set $\alpha$ to 2 and $\lambda_{PT}$ to 0.2. For the other datasets, we set $\alpha$ to 0.5 and $\lambda_{PT}$ to 0.1. For the second stage, we set the batch size to 512 on CIFAR-10(N) and CIFAR-100(N), 256 on ANIMAL-10N, and 64 on Clothing1M and Webvision. We used the same data augmentation strategy as SimCLR in the pre-training stage and used the stronger augmentation strategy proposed in DM-AugDesc\cite{nishi2021augmentation} in the semi-supervised noise correction step. To make a fair comparison, we used the same network architecture as comparison approaches unless otherwise noted. For CIFAR-10(N) and CIFAR-100(N), we used PreAct ResNet-18 \cite{he2016identity}. For ANIMAL-10N, we used VGG19 \cite{simonyan2014very}. For Clothing1M, a ResNet-50 \cite{he2016identity} was used, and for WebVision, we chose inception-resnet v2 \cite{szegedy2017inception}. For all the experiments, we used an SGD optimizer with momentum 0.9 and weight decay $5\times 10^{-4}$. We use a learning rate of 0.01 for most benchmarks, except for Clothing1M we use 0.002. To demonstrate the robustness of our methods, we used the same initial learning rates and optimizer scheduling policy for a given benchmark. Most of the experiments can be run on a single NVIDIA-A100 GPU. 


\setlength{\tabcolsep}{1mm}
\begin{table}
\setlength{\abovecaptionskip}{-0.5mm}
\setlength{\belowcaptionskip}{-1mm}
\begin{center}
\caption{Comparison on CIFAR-10N and CIFAR-100N. C10-N(X) denotes five different types of noisy labels on CIFAR-10N. We present the mean and confidence intervals over five runs.}
\label{table:cifar-N}
\scalebox{0.7}{
\begin{tabular}{lcccccc}
\hline\noalign{\smallskip}
\multirow{2}{*}{\textbf{Methods}} & C10-N & C10-N & C10N & C10-N & C10-N & C100-N\\
 & (Agg.) & (R1) & (R2) & (R3) & (Worst) & (Noisy) \\
\noalign{\smallskip}
\hline
\noalign{\smallskip}
CAL \cite{Zhu_2021_CVPR} & $91.97 \pm 0.32$ & $90.93 \pm 0.31$ & $90.75 \pm 0.30$ & $90.74 \pm 0.24$ & $85.36 \pm 0.16$ & $61.73 \pm 0.42$ \\
ELR$_+$ \cite{liu2020early} & $94.83 \pm 0.10$ & $94.20 \pm 0.24$ & $95.23 \pm 0.07$ & $94.34 \pm 0.22$ & $91.09 \pm 1.60$ & $66.72 \pm 0.07$ \\
PES(Semi)  & $94.66 \pm 0.18$ & $94.45 \pm 0.14$ & $95.19 \pm 0.23$ & $95.22 \pm 0.13$ & $92.68 \pm 0.22$ & $70.36 \pm 0.33$ \\
Cores \cite{cheng2020learning} & $95.25 \pm 0.09$& $95.06 \pm 0.15 $ & $94.88 \pm 0.31$ & $94.74 \pm 0.03$ & $91.66 \pm 0.09$ & $61.15 \pm 0.73$ \\
DivideMix  & $95.01 \pm 0.71$ & $95.16 \pm 0.19$ & $95.23 \pm 0.07$ & $95.21 \pm 0.14$ & $92.56 \pm 0.42$ & $71.13 \pm 0.48$ \\
\hline
\textbf{ChiMera} & $\textbf{96.30} \pm \textbf{0.10}$ & $\textbf{95.93} \pm \textbf{0.07}$ & $\textbf{96.06} \pm \textbf{0.10}$ & $\textbf{96.33} \pm \textbf{0.03}$ & $\textbf{94.35} \pm\textbf{0.15}$ & $\textbf{72.09} \pm \textbf{0.11}$
\\
\hline
\end{tabular}
}
\end{center}
\end{table}

\section{Experimental results}

\subsection{Evaluating ChiMera at various noise ratios on simulated symmetric and instance dependent noisy labels}

We first sought to evaluate the performance of our framework at various ratios of symmetric noise using simulated experiments on CIFAR-10 and CIFAR-100 datasets (Table \ref{table:cifar-sym}). 
We found that ChiMera achieved the best performance under all simulated noise ratios on both datasets, suggesting ChiMera is robust to a wide range of noise ratios as it fuses the advantage of contrastive learning and semi-supervised learning. We found the average performance improvements provided by ChiMera (0.94\% on CIFAR-10 and 3.65\% on CIFAR-100) are not trivial by looking at the performance difference between the best CL-based baseline REED (worse in a smaller noise ratio) and semi-supervised learning-based baseline DM-AugDesc (worse in a larger noise ratio). This suggests that MixCLR successfully boosts the power of each module as well as provides a good fusion of two objectives.
We then noticed that the improvement of
ChiMera against the best-performed baseline REED is larger in CIFAR-100 than in CIFAR-10. The larger number of classes on CIFAR-100 results in a smaller number of samples per class and is more sensitive to noisy labels. ChiMera leverages MixCLR to supplement many new class-mixed samples in the feature space to learn from the large class space more efficiently, partially alleviating the disturbance of noisy labels. 
Moreover, we also observed that the improvement of our method against DM-AugDesc, is greater with the increase of noise ratios, again confirming the importance of using mixup-enhanced contrastive learning when there are many noisy labels.

We then sought to evaluate the performance of ChiMera algorithm at various ratios of instance-dependent label noise using simulated experiments on CIFAR-10 and CIFAR-100 datasets (Table \ref{table:cifar-instance}). 

We found ChiMera achieved the best performance under all simulated noise ratios (20\% and 40\%) on both datasets. We observed constant improvement (from 0.4\% to 1.5\%) by our methods against other approaches on all four types, reassuring that our method is robust to different types of noise labels and noise ratios. We also noticed that the improvement of ChiMera versus the best-performed baseline PES(Semi) is larger on CIFAR-100 than on CIFAR-10. As discussed in section \ref{sec:dataset}, CIFAR-100 has more classes and fewer samples in each class, thus being more disturbed by noisy labels. This further suggested our model's ability to partially alleviate the disturbance from noisy labels. 
\setlength{\tabcolsep}{1pt}
\begin{table}
\setlength{\abovecaptionskip}{-0.5mm}
\setlength{\belowcaptionskip}{-1mm}
\begin{center}
\caption{Comparison on Clothing1M dataset using different pretraining strategies and mini-WebVision dataset with directly transferring the trained model to ILSVRC12 dataset.}
\label{table:clothing1m_webvision}
\scalebox{0.87}{
\begin{tabular}{lccc|lcc}
\hline\noalign{\smallskip}
\multirow{2}{*}{\textbf{Methods/Pre-Train}} & Sim & Mix & Image &\multirow{2}{*}{\textbf{Methods/Dataset}} & mini-Web & ILSV-\\
&  CLR&CLR &Net & & Vision & RC12\\
\noalign{\smallskip}
\hline
\noalign{\smallskip}
Me-Momentum \cite{bai2021me} & 72.58 & 72.75 & 73.13 & Co-teaching \cite{han2018co} &63.58 & 61.48
\\SOP~\cite{liu2022robust} & 72.55 & 72.94 & 73.51 & SOP~\cite{liu2022robust} & 76.6 & 69.1 
\\DivideMix \cite{li2020dividemix}& 74.27 & 74.41 & 74.76 &DivideMix \cite{li2020dividemix} &  77.32 & 75.20 
\\LaCol~\cite{yan2022noise} & 74.09 & 74.34 & 74.68 & MoPro \cite{li2020mopro}& 77.59 & 76.31 
\\ELR$_+$ \cite{liu2020early} & \underline{74.58} & 74.53 & 74.81 & ELR$_+$ \cite{liu2020early} & 77.78  &70.29 
\\ BLTM-V~\cite{yang2022estimating} & 72.51 & 72.87 & 73.39 & ProtoMix & 77.8 & 74.4 
\\kMEIDTM~\cite{cheng2022instance} & 73.54 & 74.07 & 74.82 & SPR~\cite{wang2022scalable} & 78.12 & -
\\ DM-AugDesc \cite{nishi2021augmentation} & 74.47 & 74.54 & 75.11  & DM-AugDesc \cite{nishi2021augmentation} & 78.64  &75.52 
\\ TO-Fliter~\cite{feng2023ot} &73.95 & 74.23 & 74.5& LongReMix \cite{cordeiro2021longremix} & 78.92  & -
\\
TCL~\cite{huang2023twin} &73.67 &74.21 & 74.79 & TCL~\cite{huang2023twin} & 79.1 & 75.4 
\\DM-CNLCU \cite{xia2021sample} & 74.24 & 74.50 & 74.91 & NGC \cite{wu2021ngc} & 79.16 & 74.44 \\PES(Semi) \cite{bai2021understanding} & 73.60 & 73.77 & 74.99 & GJS \cite{englesson2021generalized}& 79.28& 75.50
\\TSCSI\_IDN~\cite{zhao2022centrality} & 74.01 & \underline{74.65} & \textbf{75.40 } &TSCSI\_IDN~\cite{zhao2022centrality} & 79.36 & 76.08 
\\
CDLNL~\cite{cheng2022class} & 73.66 & 74.18 & 75.12  &  C2D \cite{zheltonozhskii2021contrast} & 80.21 &  \underline{76.64}
\\
DISC~\cite{li2023disc} & 74.14 & 74.38 & 74.79 & DISC~\cite{li2023disc} & 80.28 & 77.44 
\\NCR~\cite{iscen2022learning} & 74.01 & 74.21 & 74.6 & NCR~\cite{iscen2022learning} & 80.5 & - 
\\
\hline
ChiMera (w/o MixCLR) & 74.45 & 74.47 & 75.12 & ChiMera (w/o MixCLR) & \underline{80.68} & 76.43 
\\ 
\textbf{ChiMera (Ours)} & \textbf{74.67} & \textbf{74.76} & \underline{75.31} & \textbf{ChiMera (Ours)} & \textbf{80.96} &  \textbf{76.78} \\
\hline
\end{tabular}
}
\end{center}
\end{table}

\subsection{MixCLR is better than other CL methods that also use mixup augmentation}

To further prove the effectiveness of MixCLR, we conducted extensive ablation study between MixCLR and other publicly available existing methods such as MixCo~\cite{kim2020mixco}, i-Mix~\cite{lee2020mix}, Un-Mix~\cite{shen2020mix}, DACL~\cite{verma2021towards}, and M-Mix~\cite{zhang2022m} that also leverage mixup and contrastive learning, the representative of those methods and show this critical difference leads to the superior performance of MixCLR on improving label diffusion under noisy label setting (Table \ref{tab:i-mix}). We hypothesize that the consistent improvement comes from the fact that MixCLR is the only method that learns the representation of mixed samples via the self-supervision signal, while other existing similar ideas of using image mixtures in unsupervised learning are either contrasting the mixed views and original views or using different $\lambda$ when constructing mixed pairs to contrast.

\setlength{\tabcolsep}{1mm}
\begin{table}[ht]
\setlength{\abovecaptionskip}{-0.5mm}
\setlength{\belowcaptionskip}{-1mm}
\begin{center}
\caption{Comparison on CIFAR-10 with simulated 40\% asymmetric noise. }
\label{table:cifar-asym}
\begin{tabular}{lcc}
\hline\noalign{\smallskip}
Methods  &Peak Acc (\%)  & Final Acc (\%) \\
\noalign{\smallskip}
\hline
Cross Entropy & 85.00 & 72.30\\
CNLCU-H \cite{xia2021sample} & 74.93 & 73.40\\
$\text{ELR}_+$(ResNet-34) \cite{liu2020early} & 93.11 & 92.98\\
DivideMix \cite{li2020dividemix} & 93.43 & 92.17 \\
LRR \cite{li2021learning}& 93.29 & 92.45\\
DM-AugDesc \cite{nishi2021augmentation} &94.66 & 94.31 \\

OT-Filter~\cite{feng2023ot} & 95.23 & 95.08 \\
\hline
\textbf{ChiMera (Ours)}& \textbf{95.61} &\textbf{ 95.43} \\
\hline
\end{tabular}
\end{center}
\end{table}
\subsection{Real-world noisy label datasets}
After verifying the performance and robustness of ChiMera on simulated noisy label datasets, we next evaluated it on three more challenging real-world benchmarks, where the types of noisy labels are unknown and could be different from simulated noise. We used the same hyperparameters as those used in the CIFAR-10 and CIFAR-100 experiments to verify the insensitivity of our method to hyperparameters. We first compared ChiMera with other leading comparison approaches on six types of noisy labels in CIFAR-10N and CIFAR-100N (Table \ref{table:cifar-N}). We observed constant improvement (from 0.77\% to 1.79\%) by our methods against other approaches on the six types, reassuring that our method is robust to different types of noise labels and noise ratios. Moreover, our method consistently presented smaller confidence intervals, suggesting the robustness of our methods. Among the six types of noisy labels, the improvement of our method is highest in the Worst category, which is the most challenging noisy label setting that contains a diverse set of noisy labels, indicating the effectiveness of our method in handling challenging noisy labels in real-world applications.

Next, we evaluated ChiMera on two more challenging large-scale datasets Clothing1M and mini-Webvision (Table \ref{table:clothing1m_webvision}).
We found that with MixCLR pretraining, ChiMera obtained the best performance on the Clothing1M dataset if no pretraining from an extra dataset such as ImageNet \cite{krizhevsky2017imagenet} is provided. Even with the pretraining, ChiMera still achieves a comparable performance to the best-performed baseline TSCSI\_IDN \cite{zhao2022centrality}.  We then investigated the advantage of pre-training the model using MixCLR against SimCLR. Our ablation studies on Clothing1M indicate that all methods, including comparison approaches, achieved improved or comparable performance when using MixCLR at the pre-training stage. For mini-WebVision, we found that ChiMera achieves the best performance, even without the MixCLR optimization in stage II. With the ongoing MixCLR utilization, ChiMera also achieves the best transfer learning ability on the ILSVRC12 dataset. 
\setlength{\tabcolsep}{1pt}
\begin{table}[ht]
\setlength{\abovecaptionskip}{-0.5mm}
\setlength{\belowcaptionskip}{-3mm}
\begin{center}
\caption{Comparison on the asymmetric noisy label dataset ANIMAL-10N. Columns denote different pretraining strategies. }
\label{table:animals}
\begin{tabular}{lcccc}
\hline\noalign{\smallskip}
\textbf{Methods / Pretraining} & None & SimCLR & MixCLR & ImageNet\\
\noalign{\smallskip}
\hline
\noalign{\smallskip}
DISC~\cite{li2023disc} & 84.4 & 85.6 & 86.2 & 87.1 
\\SPR~\cite{wang2022scalable} & 84.3 & 84.9 & 85.5 & 86.8
\\DivideMix \cite{li2020dividemix} & 85.8 & 87.2 & 88.1 & 88.8
\\DM-AugDesc \cite{nishi2021augmentation} & 86.0 & 87.8 & 88.3 & 89.1

\\ \hline
\textbf{ChiMera} (w/o AsyMixCLR) & \underline{87.1} & \underline{88.6} & \underline{88.7} & \underline{89.3}
\\\textbf{ChiMera} & \textbf{87.5} & \textbf{88.9} & \textbf{89.2} & \textbf{89.5}\\
\hline
\end{tabular}
\end{center}
\end{table}
\vspace{-2mm}
\subsection{Addressing asymmetric noisy labels}

Finally, we studied the performance of ChiMera on asymmetric noisy labels. We considered a simulated asymmetric noise dataset based on CIFAR-10 (Table \ref{table:cifar-asym}) and a real-world asymmetric noise dataset ANIMAL-10N (Table \ref{table:animals}). The same as the observation on symmetric noise datasets, our method achieved the best performance on both asymmetric noise datasets. We studied the importance of the asymmetric noise-specific AsyMixCLR loss and observed its prominent performance in the ablation study. The other ablation studies demonstrated the importance of using MixCLR in both pre-training and semi-supervised learning stages.



\subsection{Ablation studies}
To fully understand the effectiveness of ChiMera, we performed a detailed ablation study on CIFAR-10 and CIFAR-100 (Table \ref{table:cifar-sym-ablation} and \ref{table:cifar-asym-ablation}). We first implemented a few variants of comparison approaches to understand the prominent performance of MixCLR. We first found that pre-training the model using SimCLR or MixCLR substantially improved the performance of comparison approaches by adding a pretraining stage to DivideMix, ELR$_+$(ResNet-34) and PES. We then found that MixCLR results in a larger performance gain than SimCLR in the pre-training stage when applied to DivideMix, indicating the superior performance of MixCLR against SimCLR.
\begin{table}[ht]
\setlength{\abovecaptionskip}{-0.5mm}
\setlength{\belowcaptionskip}{-1mm}
\begin{center}
\caption{Ablation studies on CIFAR-10 and CIFAR-100 by simulating symmetric noise with different noise ratios. `PT(SimCLR)' denotes using SimCLR in the pre-training stage. `PT(MixCLR)' denotes using MixCLR in the pre-training stage. `CL/SupCL' denotes using CL/SupCL loss on clean subset $\mathcal{C}$. }
\label{table:cifar-sym-ablation}
\scalebox{0.9}{
\begin{tabular}{l|llll|llll}
\hline\noalign{\smallskip}
Methods (Acc(\%))&\multicolumn{4}{c|}{CIFAR-10} & \multicolumn{4}{c}{CIFAR-100}\\
\hline
Noise Ratio & 20\% & 50\%  & 80\% & 90\% & 20\% & 50\% & 80\% & 90\%\\
\hline

Comparison approaches\\
\hline
PT(SimCLR)+$\text{ELR}_+$(R-34) &96.83 &95.96 &93.67 & 89.94 &79.18 &76.33 &64.72  &55.21\\
PT(SimCLR)+DivideMix &96.41 &95.33 &94.42  &93.36 &78.59 &76.37 &66.72  &53.56 \\
PT(MixCLR)+DivideMix & 96.55 & 95.76 & 94.71  & 93.41 & 79.54 & 77.87 & 66.89  & 53.89 \\
PT(MixCLR)+PES &96.21 &94.86 &93.23 &84.73 &78.93 &74.78 &63.04 &45.94 \\
\hline
\textbf{Variants of ChiMera }\\
\hline
PT(SimCLR)+SupCL & 96.95 & 96.47 & 94.62 & 93.46 & 81.38 & 78,29 & 68.16 & 54.14\\
PT(SimCLR)+SupCL+MixCLR & 96.80 & 96.49 & 95.42 & 93.14 & 81.21 & 79.27 & 68.42 & 54.57\\
PT(MixCLR)+CL+MixCLR & 96.71 & 96.36 & 95.12 & 93.61 & 81.01 & 78.97 & 68.29 & 56.10\\
PT(MixCLR)+SupCL & 96.92 & 96.56 & 95.10 & 93.51 & 81.37 & 79.28 & 68.33 & 56.23\\
\hline
\textbf{ChiMera (Ours)} & \textbf{96.99} & \textbf{96.57} & \textbf{95.62} & \textbf{93.71 }& \textbf{81.44} & \textbf{79.62} & \textbf{68.49}& \textbf{56.28}\\
\hline
\end{tabular}
}
\end{center}
\end{table}

We then implemented a few variants of our method to study the contribution of each component in our method. We found that using MixCLR in the pretraining stage and the semi-supervised learning stage both improve the performance. By comparing these two stages, using MixCLR in the pre-training stage results in a greater improvement. Although noisy labels are not used in the pre-training stage, MixCLR is still able to assist the warm-up stage by providing mixed-up augmentation and later ease the semi-supervised learning stage. The improvement of our method is in general larger when there are more noisy labels, again indicating the importance of using MixCLR to pre-train the model. We also found that using SupCL for symmetric noise results in a large performance gain in the semi-supervised stage, demonstrating the effectiveness of SupCL in utilizing supervised labels to obtain better representation. However, when asymmetric noise is presented, we found CL is a better option, which is reasonable as asymmetric noise usually leads to less reliable noise detection results.

\setlength{\tabcolsep}{1mm}
\begin{table}[!ht]
\setlength{\abovecaptionskip}{-0.5mm}
\setlength{\belowcaptionskip}{-2mm}
\begin{center}
\caption{Ablation studies on CIFAR-10 with 40\% asymmetric noise. `(Asy)MixCLR' denotes applying (Asy)MixCLR loss.}
\label{table:cifar-asym-ablation}
\begin{tabular}{lcc}
\hline\noalign{\smallskip}
Methods  &Peak Acc (\%)  & Final Acc (\%) \\
\noalign{\smallskip}
\hline
PT(SimCLR)+DivideMix&93.45 & 90.75 \\
PT(SimCLR)+$\text{ELR}_+$(ResNet-34) & 94.32 & 93.78 \\
\hline
\textbf{Variants of ChiMera}\\
\hline
PT(SimCLR)+SupCL &94.37 & 94.12 \\
PT(MixCLR)+SupCL+MixCLR & 94.67 & 94.49 \\
PT(SimCLR)+CL &95.35 & 95.17  \\
PT(SimCLR)+CL+MixCLR & 95.43 & 95.30 \\
PT(MixCLR)+CL+MixCLR & 95.46 & 95.30 \\
\textbf{ChiMera (Ours, w/ AsyMixCLR)}& \textbf{95.61} &\textbf{ 95.43} \\
\hline
\end{tabular}
\end{center}
\end{table}

We finally present a detailed ablation study on the choice of major hyperparameter: $\{\alpha, \tau, E_w, \lambda_{pt}, \lambda_{cl}, \lambda_{mix}, \lambda_{asym}\}$ in ChiMera on CIFAR-10 and CIFAR-100 (Table \ref{table:hyper_ablation}). We found that the core hyperparameter of MixCLR to construct mixed views ($\alpha, \tau$) and weights for pretraining ($\lambda_{pt}$) is stable. We also find that under a higher noise ratio, the choice of hyperparameters used in warming up ($E_w$) and stage II ($\delta, \lambda_{cl}$) becomes more unstable. The confidence interval also becomes larger. We observe that a smaller warmup epoch $E_w$, smaller ongoing SupCL and MixCLR weights, and a more conservative noise detection threshold (smaller $\tau$) may lead to better results when the noise ratio is high. Still, we find the warmup epoch affects the confidence interval most, suggesting the need for a better way to improve the warmup process.

\begin{table}
    \setlength{\abovecaptionskip}{-0.5mm}
    \setlength{\belowcaptionskip}{-1mm}
    \centering
    \caption{Ablation study results of different hyperparameters. }
    \begin{tabular}{l|c|cc|cc}
        \hline
        & & \multicolumn{2}{c|}{CIFAR-10} & \multicolumn{2}{c}{CIFAR-100} \\
         & Choice & 20\% & 80\% &  20\%& 80\% \\
        \hline
        \multirow{3}{*}{$\lambda_{pt}$} & 0.1 & \textbf{96.99 ± 0.12} & \textbf{95.62 ± 1.67} & \textbf{81.44 ± 0.15} & \textbf{68.49 ± 0.49} \\
                                    & 0.2 & 96.73 ± 0.11 & 95.06 ± 1.65 & 80.57 ± 0.17 & 67.01 ± 0.52 \\
                                    & 1   & 95.21 ± 0.15 & 93.23 ± 1.73 & 79.43 ± 0.19 & 66.48 ± 0.56 \\
        \hline
        \multirow{4}{*}{$\alpha$} & 0.5 & 96.81 ± 0.13 & 95.23 ± 1.61 & 81.12 ± 0.16 & 67.94 ± 0.50 \\
                                & 1   & 96.78 ± 0.17 & 95.28 ± 1.72 & 81.27 ± 0.19 & 68.12 ± 0.58 \\
                                & 2   & \textbf{96.99 ± 0.12} & \textbf{95.62 ± 1.67} & \textbf{81.44 ± 0.15} & \textbf{68.49 ± 0.49} \\
                                & 4   & 95.73 ± 0.11 & 94.21 ± 1.64 & 80.3  ± 0.14 & 67.15 ± 0.47 \\
        \hline
        \multirow{3}{*}{$\tau$} & 0.5 & \textbf{96.99 ± 0.12} & \textbf{95.62 ± 1.67} & \textbf{81.44 ± 0.15} & \textbf{68.49 ± 0.49} \\
                              & 0.1 & 95.91 ± 0.11 & 94.07 ± 1.63 & 80.69 ± 0.16 & 67.57 ± 0.48 \\
                              & 0.05 & 95.34 ± 0.13 & 93.11 ± 1.61 & 80.12 ± 0.16 & 66.89 ± 0.50 \\
        \hline
        \multirow{3}{*}{$\lambda_{cl}$} & 1 & \textbf{96.99 ± 0.12} & 95.44 ± 1.54 & \textbf{81.44 ± 0.15} &\textbf{ 68.49 ± 0.49} \\
                                    & 0.1 & 96.94 ± 0.15 & \textbf{95.62 ± 1.67} & 81.41 ± 0.18 & 68.43 ± 0.53 \\
                                    & 0.01 & 96.95 ± 0.12 & 95.21 ± 1.59 & 81.36 ± 0.15 & 68.19 ± 0.52 \\
        \hline
        \multirow{2}{*}{$\delta$} & 0.5 & \textbf{96.99 ± 0.12} & \textbf{95.62 ± 1.67} & \textbf{81.44 ± 0.15} & 68.21 ± 0.47 \\
                                   & 0.03 & 96.89 ± 0.14 & 95.41 ± 1.72 & 81.29 ± 0.14 & \textbf{68.49 ± 0.49} \\
        \hline
        \multirow{3}{*}{$\lambda_{mix}$} & 0.2 & \textbf{96.99 ± 0.12} & 95.54 ± 1.55 & \textbf{81.44 ± 0.15} & 68.39 ± 0.48 \\
                                     & 0.1 & 96.97 ± 0.13 & \textbf{95.62 ± 1.67} & 81.39 ± 0.14 & \textbf{68.49 ± 0.49} \\
                                     & 0.01 & 96.93 ± 0.15 & 95.29 ± 1.61 & 81.35 ± 0.17 & 68.34 ± 0.51 \\
        \hline
        \multirow{3}{*}{$E_w$} & 2 & 94.67 ± 1.24 & \textbf{95.62 ± 1.67} & 76.14 ± 1.43 & 66.92 ± 1.77 \\
                             & 5 & 95.36 ± 0.31 & 95.10 ± 1.08 & 79.64 ± 0.56 & 68.17 ± 1.23 \\
                             & 10 & \textbf{96.99 ± 0.12} & 94.64 ± 0.31 & \textbf{81.44 ± 0.15} & \textbf{68.49 ± 0.49} \\
        \hline

    \end{tabular}
    \label{table:hyper_ablation}
\end{table}

\section{Conclusions}
In this paper, we have studied the problem of learning with noisy labels. We have proposed ChiMera, which performs contrastive learning between mixed-up augmentations via a novel and flexible contrastive objective MixCLR. The intuition of ChiMera is to learn and refine the representation of the mixed-up samples to alleviate the disturbance from noisy labels. ChiMera has obtained state-of-the-art performance on CIFAR-10, CIFAR-100, CIFAR-10N, CIFAR-100N, Clothing1M, mini-WebVision, and Animal-10N under both symmetric and asymmetric noise settings. In the future, we plan to apply ChiMera and MixCLR to other applications and other data modalities, and further improve ChiMera by utilizing extra validation supervision.


\bibliographystyle{IEEEtran}
\bibliography{TPAMI}
\end{document}


\title{ChiMera: Learning with noisy labels by contrasting mixed-up augmentations}

\author{IEEE Publication Technology,~\IEEEmembership{Staff,~IEEE,}
\thanks{This paper was produced by the IEEE Publication Technology Group. They are in Piscataway, NJ.}
}

\markboth{IEEE TRANSACTIONS ON MULTIMEDIA ,~Vol.~14, No.~8, May~2023}%
{Shell \MakeLowercase{\textit{et al.}}: A Sample Article Using IEEEtran.cls for IEEE Journals}


\maketitle
\section{Details of MixMatch Loss}
In this section, we introduce the details of the implementation of MixMatch loss \cite{li2020dividemix} $\mathcal{L}_{MixMatch}$ used in main paper Section 2.4. 
\subsection{Dividing samples into clean and noisy sets based on the prediction loss.}
Let $\mathbf{D}_{\mathcal{N}}=\{\mathbf{x}_i\, \tilde{\mathbf{y}}_i\}_{i=1}^{|\mathbf{D}_{\mathcal{N}}|}$ be the noisy dataset and
$\theta = \{\theta_{feat}, \theta_{proj}, \theta_{cls}\}$ be the model, we first use the classifier $M_{cls}= \{f(.|\theta_{enc}), G(f(.)|\theta_{cls})\}$ to generate the predicted labels $\{\mathbf{p}_i\}_{i=1}^{|\mathbf{D}_{\mathcal{N}}|}$. We can then calculate the prediction loss (cross-entropy loss) as:
\begin{equation}
    \begin{aligned}
    \mathcal{L}_{ce,i} &=  -\sum_{c=1}^{\mathbf{C}}\tilde{y}_i^c\log p_i^c.
    \end{aligned}
    \label{eqn:ce}
\end{equation}
We then fit a two-component Gaussian Mixture Model (GMM) on
$\{\mathcal{L}_{ce,i}\}_{i=1}^{|\mathbf{D}_{\mathcal{N}}|}$ using Expectation Maximization algorithm. Here we assume the Gaussian component with smaller mean values to be the clean cluster and acquired $w_i$ as the probabilities of sample $\mathbf{x}_i$ belongs to the clean subset. We then divide the whole dataset into a clean subset $\mathcal{C}\triangleq \{\mathbf{x}_i, \tilde{\mathbf{y}}_i,w_i\}_{i=1}^{|\mathcal{C}|}$ and $\mathcal{U}\triangleq \{\mathbf{u}_i\}_{i=1}^{|\mathcal{U}|}$ by hard coding a threshold $\tau$ on $\{w_i\}_{i=1}^{|\mathbf{D}_{\mathcal{N}}|}$. 
\subsection{Guessing, refining labels and MixMatch.}
\subsubsection{Guessing and refining labels.} Given a classifier $M_{cls}$, we first obtain the prediction probabilities $\mathbf{p}_i$ of every samples $\mathbf{x}_i$. For a unlabeled sample $\mathbf{u}_i$, we use $\mathbf{p}_i$ as its predicted pseudo labels. Specifically, we utilize the consistency regularization proposed by MixMatch \cite{berthelot2019mixmatch}. Let $\mathbf{x}_{i,m}$ be the $m$-th augmentation of $\mathbf{x}_i$, $ \overline{\mathbf{p}}_{i,m}$ be the prediction of the m-th augmentation given by $M_{cls}$, we aggregate these predictions by taking the average:
\begin{equation}
    \begin{aligned}
    \overline{\mathbf{p}}_i &= \frac{1}{M}\sum_{m=1}^{M} \overline{\mathbf{p}}_{i,m}.
    \end{aligned}
    \label{eqn:average}
\end{equation}
For a labeled sample, we refine the labels by incorporating the model's prediction and clean probabilities:
\begin{equation}
    \begin{aligned}
    \mathbf{\overline{y}}_i &= w_i\tilde{\mathbf{y}}_i + (1-w_i) \overline{\mathbf{p}}_i.
    \end{aligned}
    \label{eqn:refine}
\end{equation}
We then apply sharpen operations to the generated labels (both labeled and unlabeled) with temperature $T$ to concentrate the refined labels. For $c=1,\dots,\mathbf{C}$:
\begin{equation}
    \begin{aligned}
    \hat{\mathbf{y}}^c_i &=\frac{(\overline{y}^c_{i} )^{\frac{1}{T}}}{\sum_{c=1}^{\mathbf{C}}(\overline{y}^c_{i} )^{\frac{1}{T}}},\\
    \hat{\mathbf{p}}^c_i &=\frac{(\overline{p}^c_{i} )^{\frac{1}{T}}}{\sum_{c=1}^{\mathbf{C}}(\overline{p}^c_{i} )^{\frac{1}{T}}}.
    \end{aligned}
    \label{eqn:sharpen}
\end{equation}
\subsubsection{MixMatch.} \label{sec:guess}Given a labeled minibatch $\mathcal{X}_B=\{\mathbf{x}_i, \hat{\mathbf{y}}_i\}_{i=1}^{B}$ and an unlabeled minibatch $\mathcal{U}_B=\{\mathbf{u}_i, \mathbf{p}_i\}_{i=1}^{B}$, we apply MixMatch to get mixed-up minimatches $\mathcal{X}'_B$ and $\mathcal{U}'_B$. Concretely, given a labeled sample $(\mathbf{x}_i, \hat{\mathbf{y}}_i)$ and an unlabeled sample $(\mathbf{u}_j, \hat{\mathbf{p}}_j)$, we first sample a mixup coefficient $\lambda \sim Beta(\alpha,\alpha)$. Let $\lambda_1 = \max(\lambda, 1-\lambda)$, to construct a new mixed-up labeled sample $\mathbf{x}'_{i,j}\in \mathcal{X}'_B$ and its pseudo labels $\mathbf{y}'_{i,j}$, we apply the following mixup operation \cite{berthelot2019mixmatch,zhang2017mixup}:
\begin{equation}
    \begin{aligned}
    \mathbf{x}'_{i,j}  &= \lambda_1 \mathbf{x}_i + (1-\lambda_1)\mathbf{u}_j,\\ 
    \mathbf{y}'_{i,j}  &= \lambda_1 \hat{\mathbf{y}}_i + (1-\lambda_1)\mathbf{p}_j.
    \end{aligned}
    \label{eqn:constructx}
\end{equation}
Let $\lambda_2=1-\lambda_1$, to construct a new mixed-up unlabeled sample $\mathbf{u}'_{j,i}\in \mathcal{U}'_B$ and its pseudo labels $\mathbf{p}'_{j,i}$, we apply the same operation as shown in Equation (\ref{eqn:constructx}), but replace the $\lambda_1$ with $\lambda_2$. This operation ensures the mixed-up labeled image is closer to the original labeled image $\mathbf{x}_i$ and vice versa. For convenience, we use $(\mathbf{x}', \mathbf{y}')$ and $( \mathbf{u}', \mathbf{p}')$ to denote $(\mathbf{x}_{i,j}', \mathbf{y}_{i,j}')$ and $( \mathbf{u}_{j,i}', \mathbf{p}_{j,i}')$. We then formally define the MixMatch process:
\begin{equation}
    \begin{aligned}
    \mathcal{X}',\mathcal{U}' = \text{MixMatch}(\mathcal{X},\mathcal{U}, T, M, \alpha).
    \end{aligned}
    \label{eqn:mixmatch-process}
\end{equation}
\subsubsection{Semi-supervised learning loss.}  Let $\hat{\mathbf{y}}', \hat{\mathbf{p}}'$ be the model's prediction given $\mathbf{x}'$ and $\mathbf{u}'$, we apply cross-entropy loss $\mathcal{L}_{\mathcal{X}}$ on $\mathcal{X}'_B$ and mean squared error loss on $\mathcal{U}'_B$: 
\begin{equation}
    \begin{aligned}
    \mathcal{L}_{\mathcal{X}'}  &= -\frac{1}{|\mathcal{X}'_B|}\sum\limits_{(\mathbf{x}', \mathbf{y}') \in \mathcal{X}'_B}\mathbf{y}'\log \hat{\mathbf{y}}',\\ 
    \mathcal{L}_{\mathcal{U}'}  &= -\frac{1}{|\mathcal{U}'_B|}\sum\limits_{(\mathbf{x}', \mathbf{p}') \in \mathcal{U}'_B}||\mathbf{p}' -  \hat{\mathbf{p}}'||_2^2.
    \end{aligned}
    \label{eqn:semi-lu}
\end{equation}
We also apply a regularization term proposed by \cite{arazo2019unsupervised,li2020dividemix,tanaka2018joint}. This term leverages a prior $\pi$ to regularize the model’s class level prediction across all samples in the dataset. For example, if the dataset is class-balanced, then the prior is $\pi_c=\frac{1}{\mathbf{C}},\  c = 1, \dots,\mathbf{C}$. The regularization loss is:
\begin{equation}
    \begin{aligned}
    \mathcal{L}_{reg}  &= -\sum\limits_{c}\pi_c\log(\pi_c\bigg/\frac{1}{|\mathcal{X}'|+|\mathcal{U}'|}\sum\limits_{x \in \mathcal{X}'+\mathcal{U}'}\hat{p}^c).
    \end{aligned}
    \label{eqn:semiloss}
\end{equation}
The final loss is: 
\begin{equation}
    \begin{aligned}
\mathcal{L}_{D\_MixMatch} &= \mathcal{L}_{\mathcal{X}'} + \lambda_u\mathcal{L}_{\mathcal{U}'}+\lambda_{reg}\mathcal{L}_{reg}.
    \end{aligned}
    \label{eqn:semi-all}
\end{equation}
\section{More Implementation Details}
In this section, we provided more implementation details. For all the pre-training experiments except for WebVision, we used batch size = 1024. For WebVision, we used batch size=512. For CIFAR-10, CIFAR-100 and ANIMAL-10N, the epoch of pre-training stage is 800. For Clothing1M, we trained 30 epochs. For WebVision, we trained for 80 epohcs. For the augmentation strategies, we leveraged the realization from SimCLR \cite{chen2020simple} in the pre-training stage, and follows the realization from DM-AugDesc \cite{nishi2021augmentation} in the second stage, which is to use `weak' augmentation for label guessing and use `strong' augmentation for semi-supervised learning. In the second stage, we also used `strong' augmentation for all the contrastive learning (SupCon, CL, MixCLR, AsyMixCLR).

For synthetic experiments on CIFAR-10 and CIFAR-100, we used learning rate 0.01 for all of our experiments. For CIFAR-10 experiments, we trained for 350 epochs with a divide-by-10 learning rate decay after 200 epochs. For CIFAR-10 experiments, we trained for 350 epochs with a divide-by-10 learning rate decay after 200 epochs. For CIFAR-100 experiments, we trained for 400 epochs with a divide-by-10 learning rate decay after 300 epochs. We listed the important hyper-parameters we used, and for the other hyper-parameters (Supplementary Table \ref{table:cifar-sym-hyper}) we followed the settings used in DM-AugDesc \cite{nishi2021augmentation}.   
\setlength{\tabcolsep}{0.5pt}
\begin{table}[ht]
\setlength{\belowcaptionskip}{1mm}
\begin{center}
\caption{Hyperparameter Sets used in synthetic experiments. }
\label{table:cifar-sym-hyper}
\begin{tabular}{l|cccc|c|cccc}
\hline\noalign{\smallskip}
\textbf{Hyperparameters}&\multicolumn{5}{c|}{\textbf{CIFAR-10}} & \multicolumn{4}{c}{\textbf{CIFAR-100}}\\
\hline
Noise Type & Sym & Sym   & Sym  & Sym  & Asym& Sym  & Sym  & Sym  & Sym \\
\hline
Noise Ratio & 20\% & 50\%  & 80\% & 90\% & 40\%& 20\% & 50\% & 80\% & 90\%\\
\hline
$\lambda_{cl}$ & 1 & 1 & 0.1 & 0.1 & 0.1 & 1 & 1 & 1 & 0.01\\
$\lambda_{u}$ & 0 & 25 & 25 & 50 & 0 & 25 & 150 & 150 & 150\\
Threshold $\tau$ & 0.5 & 0.5 & 0.5 & 0.03 & 0.5 & 0.5 & 0.5 & 0.03 & 0.03\\
$\lambda_{mix}$ & 0.2 & 0.2 & 0.1 & 0.1 & 0.2 & 0.2 & 0.2 & 0.1 & 0.01\\
$E_w$ & 10 & 10 & 2 & 2 & 10 & 10 & 10 & 10& 5\\

$\lambda_{AsyMix}$ & - & - &- & - & 0.2 & - & - &- & -\\

\hline
\end{tabular}
\end{center}
\end{table}

For experiments on CIFAR-10N, we used $\lambda_{cl}=1, \lambda_{mix}=0.2, \lambda_u=0, E_w=10$, GMM threshold $\tau=0.5$ for all 5 trials. For the other hyperparameters, we maintained them the same to the symmetric and asymmetric CIFAR-10 experiments. For experiments on CIFAR-100N, we used $\lambda_{cl}=1, \lambda_{mix}=0.2, \lambda_u=150, E_w=10$, GMM threshold $\tau=0.5$ and the confidence penalty regularization (Section \ref{sec:conf-penalty}) during warm up process.
For ANIMAL-10N experiments, we trained our models for 100 epoch. We used learning rate 0.01, and two divide-by-10 decay after 50 and 75 epochs. We used the listed hyperparameters: $\lambda_{cl}=1, \lambda_{mix}=0.05, \lambda_{AsyMix}=0.05, \lambda_u=0, E_w=5$, GMM threshold $\tau=0.5$. Since the noise type in ANIMAL-10N is asymmetric, we used the confidence penalty regularization during warm up process.

For Clothing1M experiments, we trained our models for 80 epochs, with learning rate 0.002 and a divide-by-10 decay after 40 epochs. We used the listed hyperparameters: $\lambda_{cl}=1, \lambda_{mix}=0.05, \lambda_u=0, E_w=1$, GMM threshold $\tau=0.5$. For WebVision experiments, we trained our models for 100 epochs, with learning rate 0.1 and a divide-by-10 decay after 50 epochs. We used the listed hyperparameters: $\lambda_{cl}=0.1, \lambda_{mix}=0.1, \lambda_u=0, E_w=1$, GMM threshold $\tau=0.5$.


\section{Full Comparison Results}
In this section, we provide full results compared with other baseline approaches on different benchmarks. Table \ref{table:cifar-sym} summarizes full comparisons with existing methods on CIFAR-10 and CIFAR-100 with symmetric noise. Additional baseline approaches are provided: Bootstrap \cite{reed2014training}, F-correction \cite{patrini2017making}, Co-teaching+ \cite{yu2019does}, Mixup \cite{zhang2017mixup}, P-correction \cite{yi2019probabilistic}, Meta-Learning \cite{li2019learning}, M-correction \cite{arazo2019unsupervised}, MOIT+ \cite{ortego2021multi}, Sel-CL+ \cite{li2022selec}, ProtoMix \cite{li2021learning}, and NGC \cite{wu2021ngc}. We directly report their results. 

Table \ref{table:cifar-N-S} summarizes full comparisons with existing methods on CIFAR-10N and CIFAR-100N. Additional baseline approaches are provided: Negative-LS \cite{wei2021understanding}, JoCoR \cite{wei2020combating}, and Co-teaching \cite{han2018co}. We directly report their results. 

Table \ref{table:clothing1m} summarizes full comparisons with existing methods on Clothing1M. Additional baseline approaches are provided: F-correction \cite{patrini2017making}, M-correction \cite{arazo2019unsupervised}, Joint-Optim \cite{tanaka2018joint}, Meta-Cleaner \cite{zhang2019metacleaner}, Meta-Learning \cite{li2019learning}, P-correction \cite{yi2019probabilistic}, BLTM \cite{yang2021estimating}, CAL \cite{Zhu_2021_CVPR}, VolMinNet \cite{li2021provably}, ILFC \cite{berthon2021confidence}. We directly report their results. 

Table \ref{table:webvision} summarizes full comparisons between ChiMera and other existing methods on mini-WebVision. Additional baseline approaches are provided: F-correction \cite{patrini2017making}, Decoupling \cite{malach2017decoupling}, D2L \cite{ma2018dimensionality}, MentorNet \cite{jiang2018mentornet}, Co-teaching \cite{han2018co}, Iterative-CV \cite{chen2019understanding}, ProtoMix \cite{li2021learning}, MoPro \cite{li2020mopro}, and NGC \cite{wu2021ngc}. We directly report their results. 

Table \ref{table:cifar-asym} summarizes full comparisons with existing methods on CIFAR-10 with 40\% asymmetric noise. Additional baseline approaches are provided: F-correction \cite{patrini2017making}, M-correction \cite{arazo2019unsupervised}, Iterative-CV \cite{chen2019understanding}, P-correction \cite{yi2019probabilistic}, Joint-Optim \cite{tanaka2018joint},  Meta-Learning \cite{li2019learning}, PENCIL \cite{yi2019probabilistic}, REED \cite{zhang2020decoupling}, MOIT+ \cite{ortego2021multi}, Sel-CL+ \cite{li2022selec}, ProtoMix \cite{li2021learning}, and NGC \cite{wu2021ngc}. We directly report their results. 

Table \ref{table:animals} summarizes full comparisons with existing methods on ANIMAL-10N. Additional baseline approaches are provided: SELFIE \cite{song2019selfie}, PLC \cite{zhang2021learning}, and Nested Co-teaching \cite{chen2021boosting}. We directly report their results. 

\setlength{\tabcolsep}{0.5pt}
\begin{table}[ht]
\setlength{\belowcaptionskip}{1mm}
\begin{center}
\caption{Full comparison on CIFAR-10 and CIFAR-100 by simulating symmetric noise with different noise ratios. `PT(SimCLR)' denotes using SimCLR in the pre-training stage. `PT(MixCLR)' denotes using MixCLR in the pre-training stage. `Semi' denotes the classification method in the semi-supervised learning stage. `Sup' denotes using SupCon loss on clean subset $\mathcal{C}$. }
\label{table:cifar-sym}
\resizebox{\textwidth}{52mm}{
\begin{tabular}{l|llll|llll}
\hline\noalign{\smallskip}
Methods(Acc(\%))&\multicolumn{4}{c|}{CIFAR-10} & \multicolumn{4}{c}{CIFAR-100}\\
\hline
Noise Ratio & 20\% & 50\%  & 80\% & 90\% & 20\% & 50\% & 80\% & 90\%\\
\hline
Comparison approaches\\
\hline
Bootstrap & 86.8 & 79.8 & 63.3 & 42.9 & 62.1 & 46.6 & 19.9 & 10.2 \\
F-correction & 86.8 & 79.8 & 63.3 & 42.9 & 61.5 & 46.6 & 19.9 & 10.2 \\
Co-teaching+ & 89.5 & 85.7 & 67.4 & 47.9 & 65.6 & 51.8 & 27.9 & 13.7 \\
Mixup & 95.6 & 87.1 & 71.6 & 52.2 & 67.8 & 57.3 & 30.8 & 14.6 \\
P-correction & 92.4 & 89.1 & 77.5 & 58.9 & 69.4 & 57.5 & 31.1 & 15.3 \\
Meta-Learning & 92.9 & 89.3 & 77.4 & 58.7 & 68.5 & 59.2 & 42.4 & 19.5 \\
M-correction & 94.0 & 92.0 & 86.8 & 69.1 & 73.9 & 66.1 & 48.2 & 24.3 \\
MOIT+ & 94.08 & - & 75.83 & - & 75.89 & - & 51.36 & - \\
ProtoMix & 95.8 & 94.3 & 92.4 & 75.0 & 79.1 & 74.8 & 57.7 & 29.3 \\
NGC & 95.88 & 94.54 & 91.59 & 80.46 & 79.31 & 75.91 & 62.70 & 29.76 \\
Sel-CL+ & 95.5 & 93.9 & 89.2 & 81.9 & 76.5 & 72.4 & 59.6 & 48.8 \\
\hline
Re-implemented Approaches\\
\hline
Cross-Entropy & 86.84 & 79.41 & 62.92 & 42.74 & 62.02 & 46.73 & 19.89 & 10.11\\
$\text{ELR}_+$(ResNet-34) & 95.83 & 94.88 & 93.31 & 78.73 & 77.61 & 73.60 & 60.82  & 33.41\\
DivideMix & 96.13 & 94.63 & 93.24  & 76.01 & 77.28 & 74.58 & 60.26  & 31.53 \\
Semi(PES) & 95.89 & 94.52 & 93.14 & 71.47 & 77.43 & 74.31 & 61.57 & 39.47 \\
REED & 95.78 & 95.42 & 94.36 & 93.57 & 77.03 & 72.78 & 65.64 & 54.31 \\
LRR & 95.85 & 94.57 & 92.39 & 90.17 & 79.43 & 75.01 & 65.47 & 51.32 \\
DM-AugDesc & 96.31 & 95.39 & 93.77  & 91.89 & 79.48 & 77.23 & 66.43  & 41.27 \\
\hline
Variants of comparison approaches\\
\hline
PT(SimCLR)+$\text{ELR}_+$(ResNet-34) & 96.83 & 95.96 & 93.67 & 89.94 & 79.18 & 76.33 & 64.72  & 55.21\\
PT(SimCLR)+Semi(DivideMix) & 96.41 & 95.33 & 94.42  & 93.36 & 78.59 & 76.37 & 66.72  & 53.56 \\
PT(MixCLR)+Semi(DivideMix) & 96.55 & 95.76 & 94.71  & 93.41 & 79.54 & 77.87 & 66.89  & 53.89 \\
PT(MixCLR)+Semi(PES) & 96.21 & 94.86 & 93.23 & 84.73 & 78.93 & 74.78 & 63.04 & 45.94 \\
\hline
\textbf{ChiMera(Ours) }\\
\hline
PT(SimCLR)+Semi(Sup) & 96.95 & 96.47 & 94.62 & 93.46 & 81.38 & 78,29 & 68.16 & 54.14\\
PT(SimCLR)+Semi(Sup+MixCLR) & 96.80 & 96.49 & 95.42 & 93.14 & 81.21 & 79.27 & 68.42 & 54.57\\
PT(MixCLR)+Semi(Sup) & 96.92 & 96.56 & 95.10 & 93.51 & 81.37 & 79.28 & 68.33 & 56.23\\
\hline
\textbf{PT(MixCLR)+Semi(Sup+MixCLR)} & \textbf{96.99} & \textbf{96.57} & \textbf{95.62} & \textbf{93.71 }& \textbf{81.44} & \textbf{79.62} & \textbf{68.49}& \textbf{56.28}\\
\hline
\end{tabular}
}
\end{center}
\end{table}
 \setlength{\tabcolsep}{1pt}
 \begin{table}[!htbp]
 \setlength{\belowcaptionskip}{1mm}
 \begin{center}
 \caption{Full comparison on CIFAR-10N and CIFAR-100N. C10-N(X) denotes five different types of noisy labels on CIFAR-10N. We present the mean and confidence interval over five runs.}
 \label{table:cifar-N-S}
 \resizebox{\textwidth}{22mm}{
 \begin{tabular}{lllllll}
 \hline\noalign{\smallskip}
 \textbf{Methods} & C10-N(Agg.) & C10-N(R1) & C10N(R2) & C10-N(R3) & C10-N(Worst) & C100-N(Noisy)\\
 \noalign{\smallskip}
 \hline
 Negative-LS & $91.97 \pm 0.46$ & $90.29 \pm 0.32$ & $90.37 \pm 0.12$ & $90.13 \pm 0.19$ & $82.99 \pm 0.36$ & $58.59 \pm 0.98$ \\
 JoCoR & 91.44 $\pm$ 0.05 & 90.30 $\pm$ 0.20 & 90.21 $\pm$ 0.19 & 90.11 $\pm$ 0.21 & 83.37 $\pm$ 0.30 & 59.97 $\pm$ 0.24 \\
 Co-teaching & 91.20 $\pm$ 0.13 & 90.33 $\pm$ 0.13 & 90.30 $\pm$ 0.17 & 90.15 $\pm$ 0.18 & 83.83 $\pm$ 0.13 & 60.37 $\pm$ 0.27 \\
 CAL & $91.97 \pm 0.32$ & $90.93 \pm 0.31$ & $90.75 \pm 0.30$ & $90.74 \pm 0.24$ & $85.36 \pm 0.16$ & $61.73 \pm 0.42$ \\
 ELR$_+$ & $94.83 \pm 0.10$ & $94.20 \pm 0.24$ & $95.23 \pm 0.07$ & $94.34 \pm 0.22$ & $91.09 \pm 1.60$ & $66.72 \pm 0.07$ \\
 PES(Semi) & $94.66 \pm 0.18$ & $94.45 \pm 0.14$ & $95.19 \pm 0.23$ & $95.22 \pm 0.13$ & $92.68 \pm 0.22$ & $70.36 \pm 0.33$ \\
 Cores & $95.25 \pm 0.09$& $95.06 \pm 0.15 $ & $94.88 \pm 0.31$ & $94.74 \pm 0.03$ & $91.66 \pm 0.09$ & $61.15 \pm 0.73$ \\
 DivideMix & $95.01 \pm 0.71$ & $95.16 \pm 0.19$ & $95.23 \pm 0.07$ & $95.21 \pm 0.14$ & $92.56 \pm 0.42$ & $71.13 \pm 0.48$ \\
 \hline
 \textbf{ChiMera} & $\textbf{96.30} \pm \textbf{0.10}$ & $\textbf{95.93} \pm \textbf{0.07}$ & $\textbf{96.06} \pm \textbf{0.10}$ & $\textbf{96.33} \pm \textbf{0.03}$ & $\textbf{94.35} \pm\textbf{0.15}$ & $\textbf{72.09} \pm \textbf{0.11}$\\
 \hline
 \end{tabular}
 }
 \end{center}
 \end{table}

\setlength{\tabcolsep}{1pt}
\begin{table}
\begin{center}
\caption{Full comparison on Clothing1M. Columns (rows) denotes different pre-training (semi-supervised learning) strategies. Notations are similar to the ones in Supplementary Table \ref{table:cifar-sym}.}
\label{table:clothing1m}
\begin{tabular}{lccc}
\hline\noalign{\smallskip}
\multirow{2}{*}{\textbf{Methods/Pre-Train}} & Sim & Mix & Image \\
&  CLR&CLR &Net  \\
\hline
\noalign{\smallskip}
F-correction & - & - & 69.84 \\
M-correction & - & - & 71.00 \\
Joint-Optim & - & - & 72.16 \\
VolMinNet & - & - & 72.42 \\
Meta-Cleaner & - & - & 72.50 \\
BLTM & - & - & 73.33 \\
ILFC & - & - & 73.35 \\
Meta-Learning & - & - & 73.47 \\
P-correction & - & - & 73.49 \\
CAL & - & - & 74.17 \\
DivideMix & 74.27 & 74.41 & 74.76 
\\ELR$_+$ & 74.58 & 74.53 & 74.81 
\\DM-CNLCU & 74.24 & 74.50 & 74.91 
\\PES(Semi) & 73.60 & 73.77 & 74.99 
\\DM-AugDesc & 74.47 & 74.54 & 75.11 
\\Me-Momentum(Noisy) & 72.58 & 72.75 & 73.13 
\\ Me-Momentum(Clean) & 74.93 & 75.12 & 75.18 
\\\textbf{ChiMera[Semi(Sup+MixCLR)]} & \textbf{74.67} & \textbf{74.76} & \textbf{75.31} \\
\hline
\end{tabular}
\end{center}
\end{table}

\setlength{\tabcolsep}{1pt}
\begin{table}
\begin{center}
\caption{Full comparison on mini-WebVision. Notations are similar to the ones in Supplementary Table \ref{table:cifar-sym}.}
\label{table:webvision}
\begin{tabular}{lccc|lcc}
\hline\noalign{\smallskip}
\textbf{Methods} & mini-WebVision & ILSVRC12\\
\noalign{\smallskip}
\hline
\noalign{\smallskip}
F-correction & 61.12 & 57.36 \\
D2L & 62.68 & 57.80 \\
MentorNet & 63.00 & 57.80 \\
Decoupling & 62.54 & 58.26 \\
Co-teaching & 63.58 & 61.48 \\
Iterative-CV & 65.24 & 61.60 \\
ProtoMix & 76.3 & 73.3 \\
ELR$_+$ & 77.78  &70.29 \\
MoPro & 77.59 & 76.31 \\
DivideMix &  77.32 & 75.20 \\
DM-AugDesc & 78.64  &75.52 \\
LongReMix & 78.92  & - \\
NGC & 79.16 & 74.44 \\
GJS & 79.28& 75.50
\\PT(SimCLR)+DivideMix & 80.21 &  76.64 
\\\textbf{ChiMera[Semi(Sup)]} & \textbf{80.68}  & \textbf{76.43}
\\\textbf{ChiMera[Semi(Sup+MixCLR)]} & \textbf{80.96} &  \textbf{76.78} \\
\hline
\end{tabular}
\end{center}
\end{table}

\setlength{\tabcolsep}{1mm}
\begin{table}[ht]
\setlength{\belowcaptionskip}{1mm}
\begin{center}
\caption{Full comparison on CIFAR-10 with 40\% asymmetric noise. `CL' denotes applying CL to the clean subset $\mathcal{C}$, `AsyMixCLR' denotes applying AsyMixCLR loss. Other notations are similar to the ones in Supplementary Table \ref{table:cifar-sym}.}
\label{table:cifar-asym}
\resizebox{\textwidth}{52mm}{
\begin{tabular}{lcc}
\hline\noalign{\smallskip}
Methods  &Peak Acc(\%)  & Final Acc(\%) \\
\noalign{\smallskip}
\hline
Cross Entropy & 85.00 & 72.30\\
CNLCU-H & 74.93 & 73.40\\
F-correction & 87.2 & 83.1 \\
M-correction & 87.4 & 86.3 \\
P-correction & 88.5 & 88.1 \\
Iterative-CV & 88.6 & 88.0 \\
Joint-Optim & 88.9 & 88.4 \\
Meta-Learning & 89.2 & 88.6 \\
NGC & 90.55 & - \\
PENCIL & 91.2 & - \\
ProtoMix & 91.9 & - \\
REED & 92.4 & 92.3 \\
$\text{ELR}_+$(ResNet-34) & 93.11 & 92.98\\
LRR & 93.29 & 92.45\\
MOIT+ & 93.3 & - \\
Sel-CL+ & 93.4 & - \\
DivideMix & 93.43 & 92.17 \\
DM-AugDesc &94.66 & 94.31 \\
\hline
PT(SimCLR)+DivideMix&93.45 & 90.75 \\
PT(SimCLR)+$\text{ELR}_+$(ResNet-34) & 94.32 & 93.78 \\
\hline
\textbf{ChiMera(Ours)}\\
\hline
PT(SimCLR)+Semi(Sup) &94.37 & 94.12 \\
PT(SimCLR)+Semi(CL) &95.35 & 95.17  \\
PT(SimCLR)+Semi(CL+MixCLR) & 95.43 & 95.30 \\
PT(MixCLR)+Semi(CL+MixCLR) & 95.46 & 95.30 \\
\textbf{PT(MixCLR)+Semi(CL+MixCLR+AsyMixCLR)}& \textbf{95.61} &\textbf{ 95.43} \\
\noalign{\smallskip}
\hline
\end{tabular}
}
\end{center}
\end{table}

\setlength{\tabcolsep}{1.2pt}
\begin{table}[ht]
\setlength{\belowcaptionskip}{3mm}
\begin{center}
\caption{Full comparison on the asymmetric noisy label dataset ANIMAL-10N. Columns (rows) denotes different pre-training (semi-supervised learning) strategies. Notations are similar to the ones in Supplementary Table \ref{table:cifar-asym}.}
\label{table:animals}
\begin{tabular}{lcccc}
\hline\noalign{\smallskip}
\multirow{2}{*}{\textbf{Semi-supervised learning / Pre-training}} & \multirow{2}{*}{None} & Sim & Mix & Image\\
& & CLR & CLR & Net\\
\noalign{\smallskip}
\hline
\noalign{\smallskip}
SELFIE & - & - & - & 81.8 \\
PLC & - & - & - & 83.4 \\
Nested Co-teaching & - & - & - & 84.1 \\
DivideMix & 85.8 & 87.2 & 88.1 & 88.8
\\DM-AugDesc & 86.0 & 87.8 & 88.3 & 89.1
\\\textbf{ChiMera[Semi(CL+MixCLR)]} & 87.1 & 88.6 & 88.7 & \textbf{89.3}
\\\textbf{ChiMera[(Semi(CL+MixCLR+AsyMixCLR)]} & \textbf{87.5} & \textbf{88.9} & \textbf{89.2} & \textbf{89.5}\\
\hline
\end{tabular}
\end{center}
\end{table}
\clearpage
\bibliographystyle{IEEEtran}
\bibliography{IEEE-Transactions-LaTeX2e-templates-and-instructions/egbib}